\documentclass[aps,preprint,nofootinbib,a4paper,longbibliography,eqsecnum,superscriptaddress,11pt]{revtex4-1}
\usepackage{color}

\usepackage[
	  pagebackref=false,
	  colorlinks=true,
      linkcolor=blue,
      urlcolor=blue,
      filecolor=black,
      citecolor=red,
      pdfstartview=FitV,
      pdftitle={},
        pdfauthor={},
        pdfsubject={},
        pdfkeywords={},
        pdfpagemode=None,
        bookmarksopen=true
      ]{hyperref}

\usepackage{amsmath} 
\usepackage{graphicx} 
\usepackage[space]{grffile}
\usepackage{amsthm}
\usepackage{amssymb} 
\usepackage{dsfont}
\usepackage{yfonts}
\usepackage{floatrow,dcolumn,rotating}
\usepackage{array,xcolor,graphicx}
\usepackage{booktabs,multirow}
\usepackage[utf8]{inputenc}
\usepackage{mathtools}
\usepackage{braket}
\usepackage{comment}

\usepackage[normalem]{ulem}

\usepackage{setspace}

\usepackage[all]{xy}
\usepackage{tikz}
\usetikzlibrary{arrows.meta}

\newcommand{\be}{\begin{equation}}
\newcommand{\ee}{\end{equation}}
\newcommand{\bega}{\begin{gather}}
\newcommand{\bea}{\begin{eqnarray}}
\newcommand{\eea}{\end{eqnarray}}

\newcommand{\nn}{\nonumber\\}

\newcommand{\bi}{\begin{itemize}}
\newcommand{\ei}{\end{itemize}}
\newcommand{\ben}{\begin{enumerate}}
\newcommand{\een}{\end{enumerate}}
\newcommand{\bma}{\le(\begin{matrix}}
\newcommand{\ema}{\end{matrix}\ri)}

\newcommand{\CC}{\field{C}}

\newcommand{\sL}{\mathcal{L}}

\def\Tr{{\mathrm{Tr}}}

\def\vb#1{{\mathbf{#1}}}

\def\sig{{\sigma}}
\def\le{\left}
\def\ri{\right}

\def\CC{\mathcal{C}}
\def\CD{\mathcal{D}}

\def\CJ{\mathcal{J}}
\def\CK{\mathcal{K}}
\def\CL{\mathcal{L}}

\def\CT{\mathcal{T}}

\def\CP{\mathcal{P}}

\newcommand{\pt}{\partial}

\newcommand{\vx}{\mathbf{x}}

\newcommand{\vk}{\mathbf{k}}
 
\DeclareMathOperator{\erf}{erf}

\newcommand\al{{\alpha}}
\newcommand\ep{\epsilon}

\newcommand\lam{\lambda}

\newcommand\om{\omega}

\newcommand\de{{\ensuremath{{\delta}}}}

\newcommand\vp{\varphi}

\newcommand\Th{{\Theta}}

\newcommand\ov{\over}

\newcommand\half{{\ensuremath{\frac{1}{2}}}}
\newcommand\p{\ensuremath{\partial}}

\newcommand\para{{\parallel}}

\newcommand\bfe{{\mathbf{e}}}

\begin{document}

\title{Effective field theories of dissipative fluids with one-form symmetries}
 
\author{Shreya Vardhan}
\affiliation{Stanford Institute for Theoretical Physics, Stanford University, Stanford, CA 94305
}
 
\author{Sa\v{s}o Grozdanov}
\affiliation{Higgs Centre for Theoretical Physics, University of Edinburgh, \\ Edinburgh, EH8 9YL, Scotland, 
}
\affiliation{Faculty of Mathematics and Physics, University of Ljubljana, \\ Jadranska ulica 19, SI-1000 Ljubljana, Slovenia
}

\author{Samuel Leutheusser}
\affiliation{Princeton Gravity Initiative, Princeton University, Princeton NJ 08544, USA
}

\author{Hong Liu}
\affiliation{Center for Theoretical Physics, MIT, 
Cambridge, MA 02139, USA 
}

\preprint{MIT-CTP/5752}
\begin{abstract}
\vspace{1cm}
A  system with a one-form global symmetry at finite temperature can be viewed as a dissipative fluid of string-like objects. In this work, we classify and construct the most general effective field theories for hydrodynamics of such string fluids, in a probe limit where the one-form charge density is decoupled from the energy-momentum tensor.  We show that at leading order in the derivative expansion, there are two distinct types of diffusive transport in a string fluid depending on the discrete spacetime symmetries present in it. One particular application of interest is magnetohydrodynamics (MHD), where the effective field theories describe the diffusion of magnetic field lines. Due to the distinction between the effective field theories for different discrete symmetries, we show that the MHD of a fluid with charge conjugation symmetry is qualitatively different from that of a neutron star, which we previously discussed in~\cite{Vardhan:2022wxz}. The explicit effective actions that we write down can be used to obtain the dispersion relations $\omega(k)$ up to cubic order in momenta for each of the different discrete symmetry choices. As another application of this formalism, we show that when the one-form symmetry is spontaneously broken, the effective action reduces to the Maxwell theory. This confirms the interpretation of the photon as a Goldstone boson arising from the spontaneous breaking of a one-form symmetry. 
\end{abstract}

\maketitle
\begingroup
\hypersetup{linkcolor=black}
\begin{spacing}{1}
\tableofcontents
\end{spacing}
\endgroup

\section{Introduction}\label{sec:Intro}

At large distance and time scales, the dynamics of a quantum many-body system are governed by conserved quantities and have a universal description in terms of hydrodynamics.
For conserved quantities associated with conventional ``zero-form'' global symmetries, the formulation of hydrodynamics is well-understood, and governs phenomena such as charge and momentum diffusion and sound wave propagation. The purpose of this paper is to systematically study novel hydrodynamic behaviors that arise in the presence of a one-form global symmetry~\cite{Gaiotto:2014kfa}, using the effective action formalism for hydrodynamics developed in~\cite{Crossley:2015evo, Glorioso:2017fpd}.

The notion of a higher-form global symmetry was first formalized in~\cite{Gaiotto:2014kfa}. A $p$-form symmetry acts on operators supported on $p$-dimensional manifolds, generalizing the standard case of a zero-form symmetry that acts on single points in spacetime.  Similar to continuous zero-form  symmetries, continuous higher-form symmetries are associated with conserved currents. In particular, in the case of a $U(1)$ one-form symmetry, we have the following conservation law for some two-form current $J$:  
\begin{equation} 
\pt_{\mu} J^{\mu \nu} = 0, \quad J^{\mu \nu} = - J^{\nu \mu}  . 
\label{current_cons} 
\end{equation}  
This conservation law implies that for any two-dimensional manifold $\Sigma$ in the spacetime, we can define an associated total charge 
\be 
Q_{\Sigma} = \int_{\Sigma}  \star J , \label{qdef}
\ee
which is unchanged on deforming $\Sigma$ in any direction. A system with a one-form global symmetry at finite temperature may be viewed as a fluid of one-dimensional objects, i.e., a string fluid. 

Recently, a systematic approach to hydrodynamics based on an effective action formalism was developed in~\cite{Crossley:2015evo,Glorioso:2017fpd} (see~\cite{Glorioso:2018wxw} for a review and also Refs.~\cite{Grozdanov:2013dba,Haehl:2015uoc,Jensen:2017kzi}). Starting from a current conservation law, this approach involves writing down an effective action in terms of certain collective fields associated with the conservation law that are ``integrated in'' using a Stueckelberg trick. Ref.~\cite{Crossley:2015evo, Glorioso:2017fpd} explain how to implement physical conditions such as unitarity of the microscopic dynamics and local equilibration as constraints on the effective action. The most general action consistent with these constraints can be written down in a derivative expansion, leading to constitutive relations for the conserved currents. 

In this paper, we will apply this effective field theory approach to a string fluid, starting with the conservation law \eqref{current_cons}. One physical application of particular interest is the dynamics of a strongly interacting electromagnetic plasma at long distances and late times, a regime known as ``magnetohydrodynamics'' (MHD). The one-form symmetry in this case immediately follows from the Bianchi identity, which can be written in the form 
\be 
\partial_{\mu} J^{\mu \nu} =0, \quad J^{\mu\nu} = \tilde F^{\mu \nu} = \frac{1}{2} \epsilon^{\mu \nu \rho \sigma} F_{\rho \sigma}, \label{qed}
\ee 
where $F_{\rho \sigma}$ is the electromagnetic field tensor.  The conserved quantity \eqref{qdef} in this case is the total magnetic flux passing through a surface $\Sigma$.  Previous works, starting with~\cite{Grozdanov:2016tdf}, have used this one-form symmetry to give a powerful new formulation for studying magnetohydrodynamics (MHD) with strong magnetic fields.~(See also Refs.~\cite{Hernandez:2017mch,Grozdanov:2017kyl,Grozdanov:2018fic,Glorioso:2018kcp,Armas:2018zbe,Gralla:2018kif,Benenowski:2019ule,Iqbal:2020lrt,Landry:2021kko,Armas:2022wvb,Das:2022fho,Das:2023nwl,Frangi:2024mer,Frangi:2024enh} for various developments, and in particular~\cite{Landry:2022nog} for an alternative approach to MHD with strong magnetic fields, including systems with chiral matter and Adler-Bell-Jackiw (ABJ) anomaly. More broadly, the study of string fluids was first initiated in~\cite{emparan} and~\cite{emparan_2}, and the relation of such fluids to magnetohydrodynamics was first identified in~\cite{daniel}.) Ref.~\cite{Grozdanov:2016tdf} used the approach of writing down general constitutive relations for the conserved currents in magnetohydrodynamics, and imposing various phenomenological constraints on these relations, under the assumption that all discrete spacetime symmetries (parity, charge-conjugation, and time reversal) are conserved. 

One advantage of the EFT formalism used in the present paper is that it provides a systematic derivation of the constitutive relations  in a regime with strong magnetic field, without having to impose phenomenological constraints. Moreover, it allows us to work in a general setting where we consider all possible patterns of explicit breaking of  discrete spacetime symmetries.  We work in the probe limit where the conserved two-form current does not couple to the energy-momentum tensor. As earlier emphasized in~\cite{Glorioso:2017lcn} for hydrodynamics of zero-form symmetries, the presence of different discrete spacetime symmetries in the system can lead to significantly different constitutive relations. We find that this is also the case for the hydrodynamics of string fluids. For the eleven different choices of discrete symmetry patterns (see \eqref{list_1}--\eqref{list_2} for an explicit list), the effective actions at leading order fall into two classes. These two classes have  distinct structures of the diffusive modes and dispersion relations $\omega(k)$ at $O(k^2)$.~\footnote{A partial discussion of the effect of certain discrete symmetries on the  hydrodynamics of string fluids was previously presented in~\cite{Armas:2018zbe}.}

More explicitly, the eight cases involving charge conjugation symmetry fall into one class, while the three cases without charge conjugation symmetry fall into another class where the leading-order effective action has an additional term (see \eqref{d1cases}--\eqref{d2cases}).  By analyzing the real-time evolution of various initial configurations of the one-form charge density (i.e., the magnetic field line density), we point out sharp qualitative differences in the transport behavior between the two cases. 
One system in nature where the second class is realized is a neutron star, which at the energies relevant for MHD has broken $\CC$ and $\CP$ and preserved $\CT$. In the earlier paper~\cite{Vardhan:2022wxz}, we stated the effective action for this case and discussed its physical consequences, comparing to previous phenomenological models for neutron stars such as~\cite{reis}. The remaining discrete symmetry cases correspond to MHD in exotic systems which may arise in condensed matter setups such as Weyl semimetals (see~\cite{weyl_review} for a review). While the leading diffusion behaviour always falls into one of two classes, we also derive the hydrodynamic effective actions at the next-to-leading order and show that they further distinguish the discrete symmetry choices.

One can also consider string fluids beyond MHD with a $U(1)$ one-form symmetry, where the two-form current $J^{\mu \nu}$ has a more general interpretation than \eqref{qed}. Such string fluids can be classified by considering  other ways in which the two-form current $J^{\mu\nu}$ can transform under parity and time-reversal. For example, while in MHD the action of parity and time reversal is given by 
\bea\label{01}
\CP :&&\quad (J_{0i} , J_{ij} ) \to ( J_{0i} , - J_{ij} )  \label{P-def}   , \\
\CT :&&\quad (J_{0i} , J_{ij} ) \to ( - J_{0i} ,  J_{ij} )  \label{T-def}   , 
\label{02}
\eea
we may also have a different system where $\CP, \CT$ act as  
\bea\label{03}
\CP :&&\quad (J_{0i} , J_{ij} ) \to ( - J_{0i} , J_{ij} ) \label{P+} , \\
\CT :&&\quad (J_{0i} , J_{ij} ) \to ( J_{0i} , -J_{ij} ) \label{T+} .
\label{04}
\eea
We will refer to~\eqref{01}--\eqref{02}  as $\CP_-$ and $\CT_-$, and~\eqref{03}--\eqref{04} 
 as $\CP_+$ and $\CT_+$. We may also consider  any other combination of $\CP_{\pm}$ and  $\CT_\pm$. We will show that the effective action for each of these other choices maps to one of the cases of the MHD effective action. This statement holds to all orders in the derivative expansion. As an example, the effective action for a system where the only discrete spacetime symmetry is $\CP_+\CT_-$ is the same as the one where the only discrete spacetime symmetry is $\CC \CP_-\CT_-$. The latter corresponds to the MHD effective action where the only discrete spacetime symmetry is $\CC \CP\CT$.


The EFT formalism also provides a natural way of writing down the effective action associated with the one-form symmetry \eqref{qed} when it is spontaneously broken. In this case, the effective action turns out to be the Maxwell action, and the dynamical field in the effective action can be interpreted as the photon. This provides a precise realization of the idea that the photon is the Goldstone boson associated with the spontaneous breaking of the symmetry \eqref{qed}, which has been pointed out in various references starting with~\cite{Gaiotto:2014kfa}.

The plan of the paper is as follows. In Section \ref{sec:FieldsSymm}, we describe the various inputs and constraints needed to formulate the effective field theory for hydrodynamics of a one-form symmetry. In Section \ref{sec:actions}, we write down the resulting effective actions, for which the detailed derivation is provided in Appendix~\ref{App:KMS}. In Section~\ref{sec:EFT-Sym}, we discuss concrete consequences of these effective actions in terms of novel diffusion behaviours. In Section \ref{sec:weight_three}, we discuss terms in the effective actions that lead to corrections to the diffusive dispersion relations at cubic order in momentum. In Section \ref{sec:Maxwell}, we discuss the effective action in the phase where the one-form symmetry is spontaneously broken.

\section{Fields and symmetries in fluids with a one-form symmetry}\label{sec:FieldsSymm}  

In this section, we discuss the ingredients required to construct effective field theories of dissipative fluids with a one-form symmetry. In Section \ref{sec:FieldsQS}, we discuss the field content of such theories and the way one-form symmetries constrain the structure of effective actions. In Section \ref{sec:Symm}, we then discuss various other symmetries and constraints on the effective theories, such as microscopic discrete symmetries and unitarity constraints. 

\subsection{Fields and $n$-form symmetries}\label{sec:FieldsQS}

Consider a system with a one-form symmetry at a finite temperature. 
The generating function for real-time correlators of the conserved two-form current $J^{\mu\nu}$ along a closed time path (CTP)
is given by 
\begin{align}\label{W1-1form}
e^{W[b_{1\mu\nu},b_{2\mu\nu}]} = \Tr\left[ \rho \, P e^{i \int d^4 x\,  b_{1\mu\nu} J^{\mu\nu}_1 - i \int d^4 x \, b_{2\mu\nu} J^{\mu\nu}_2 }   \right],
\end{align} 
where $b_{1\mu \nu}$ and $b_{2 \mu \nu}$ are respectively sources for $J_{\mu \nu}$ along two legs of the CTP.  Due to the current conservation law \eqref{current_cons}, 
the generating functional is invariant under a two-form gauge transformation 
\begin{align}
\label{yue}
W[b_{1\mu\nu}, b_{2\mu\nu}] = W[b_{1\mu\nu}+\partial_\mu\lambda_{1\nu} - \partial_\nu\lambda_{1\mu}  , b_{2\mu\nu}+\partial_\mu\lambda_{2\nu} - \partial_\nu\lambda_{2\mu} ] . 
\end{align}
As in the discussion of~\cite{Crossley:2015evo}, the hydrodynamic modes associated with $J^{\mu \nu}$ 
are given by the Stueckelberg fields for the local transformations of~\eqref{yue}, which are one-form fields. We will denote them 
as $A_\mu$. Note that throughout this paper, we work in 3+1 spacetime dimensions.

Using $A_\mu$ we can  express Eq.~\eqref{W1-1form} as
\begin{align}\label{W2-1form}
e^{W[b_{1\mu\nu},b_{2\mu\nu}]} = \int D A_{1\mu}  DA_{2\mu} \, e^{i S_{\rm EFT} [G_{1\mu\nu},G_{2\mu\nu}] } ,
\end{align}
where $S_{\rm EFT}$ is the effective action of $A_{s\mu}, s =1,2$ and depends only on the combinations  
\begin{align}\label{BDef}
G_{s\mu\nu} \equiv b_{s\mu\nu} + \partial_\mu A_{s\nu} - \partial_\nu A_{s\mu}   .
\end{align}
This form of the effective action ensures that \eqref{yue} is satisfied.
By definition, $G_{s\mu \nu}$ and thus $S_{\rm EFT}$ are invariant under the following transformations
\begin{align}\label{1formTconst}
A_{s \mu} \rightarrow A_{s \mu} + \pt_{\mu} \alpha_{s}, 
\end{align}
for arbitrary scalar functions $\alpha_s$. Using~\eqref{1formTconst} we can set $A_{s0} = 0$.

It is useful to introduce symmetric and antisymmetric combinations of all fields  $\varphi$ including $J_{\mu \nu}$, $A_{\mu}$, $b_{\mu \nu}$ and $G_{\mu \nu}$,
\be 
\varphi_r = \frac{\varphi_1 + \varphi_2}{2}, \quad \varphi_a = \varphi_1 - \varphi_2,
\ee
and write the effective action in terms of these ``$a$'' and ``$r$'' fields. In particular, we can obtain expressions for the current in terms of the dynamical variables as:
\be 
J_{r\mu \nu} =\frac{\delta S_{\rm EFT}}{\delta b_{a \mu \nu}} , \quad J_{a\mu \nu} =\frac{\delta S_{\rm EFT}}{\delta b_{r \mu \nu}}  .  \label{jmn}
\ee
The equations of motion for $A_{a \mu}$ and $A_{r \mu}$ then immediately imply the current conservation equation 
\be 
\partial_{\mu} J_{r, a}^{\mu \nu} = 0 . 
\ee
$J_{r\mu \nu}$ can be seen as the expectation value of the current averaged over statistical or quantum noise, and $J_{a\mu \nu}$ can be interpreted as the contribution from such noises. In the discussion below, when $J_{\mu \nu}$ appears without an $a$ or $r$ subscript, it refers to $J_{r\mu \nu}$, and we will be mostly interested in the equations of motion for these expectation values. 

To describe the fluid phase where the one-form symmetry is not broken, we require the action to be invariant under the following transformation:
\begin{equation}\label{1formT} 
A_{si} (t, \vx) \rightarrow A_{si}(t, \vx) + \lambda_i(\vx) , \quad s = 1, 2  , 
\end{equation} 
which can be interpreted as the input that each fluid element can transform independently under the one-form global symmetry. This is the natural analog in the one-form case of the ``diagonal gauge symmetry'' which was introduced to describe dissipative hydrodynamics in the zero-form symmetry case in~\cite{Crossley:2015evo}, and justified in detail in that reference.~\footnote{It would be interesting to understand whether the formalism developed in~\cite{2group} can provide a further justification for this diagonal gauge symmetry.} 
Requiring invariance of the action under \eqref{1formT} is equivalent to requiring the following symmetry transformation of the $G_{r\mu \nu}$ and  $G_{a\mu \nu}$ fields:  
\begin{align}
G_{r0i}(t,\vx) &\to G_{r0i}(t,\vx) , \label{1formT-B1} \\
G_{rij}(t,\vx) &\to G_{rij} (t,\vx) + \partial_i \lambda_j (\vx) - \partial_j \lambda_i (\vx), \label{1formT-B2}\\
G_{a\mu\nu}(t,\vx) &\to G_{a\mu\nu}(t,\vx). \label{1formT-B3}
\end{align}
Due to \eqref{1formT-B2}, in the symmetry-preserving phase, $G_{rij}$ can only appear in the effective action through the expressions $\pt_0 G_{rij}$ and $H_{ijk}$, where $H = \mathrm{d} G = \mathrm{d} b$ (cf. Eq.~\eqref{BDef}), i.e. 
\begin{align}\label{HDef}
H_{\mu \nu \lambda} = \partial_{\mu} G_{ \nu \lambda} + \partial_{\lambda} G_{\mu \nu} + \partial_{\nu} G_{\lambda\mu} \,  = \partial_{\mu} b_{ \nu \lambda} + \partial_{\lambda} b_{\mu \nu} + \partial_{\nu} b_{\lambda\mu}   .
\end{align}  
A relation which we will often use later is 
\begin{align}\label{FieldRel1}
\partial_0 G_{rij} = H_{0ij} + \partial_i G_{r0j} - \partial_j G_{r0i}  . 
\end{align}
Note that in the absence of sources, the first term on the right-hand side is zero. 

We emphasize  that in general, the hydrodynamical field $A_{\mu}$ does not have any simple relation to a microscopic vector field, such as the photon $a_\mu$ in QED. $A_\mu$ should be thought of as a collective field 
describing the low energy degrees of freedom.  We will see that $A_{\mu}$ will only equal $a_\mu$ when the one-form symmetry is spontaneously broken and the photon $a_\mu$ is realized as the Goldstone boson. To describe the dynamics in the phase with spontaneous symmetry breaking, invariance under \eqref{1formT} should not be imposed.  We will analyze this case in Section \ref{sec:Maxwell}. 

Lastly, it is important to note that the conservation of the energy-momentum tensor, $\pt_{\mu} T^{\mu \nu} = 0$, is associated with additional hydrodynamic fields, which can be interpreted as spacetime coordinates defined on the fluid spacetime background~\cite{Crossley:2015evo}. In this work, we will not couple the fields $G_{s\mu\nu}$  to those fields, and work throughout within the probe charge limit. For this reason, the fluid velocity and the temperature fields will remain non-dynamical.

\subsection{Discrete symmetries, KMS relations and constraints on the effective CTP action}\label{sec:Symm}

From the fact that the theory is formulated on the CTP contour and the microscopic dynamics is unitary, we need to impose a number of additional constraints on the structure of the effective action~\cite{Crossley:2015evo}: 
\begin{enumerate}
\item 
Every term in the effective action must have at least one $G_{a\mu\nu} $ field, as we must have
\begin{equation} \label{cons1}
S[G_{r\mu\nu}, G_{a\mu\nu} = 0] = 0.
\end{equation} 
\item 
The effective action must obey the CTP reflection symmetry
\begin{equation}  \label{cons2}
S^*[G_{r\mu\nu}, G_{a\mu\nu}] = - S[G_{r\mu\nu}, -G_{a\mu\nu}].
\end{equation} 
\item
The imaginary part of the effective action must be non-negative,
\begin{equation}  \label{cons3}
\text{Im}\, S \geq 0.
\end{equation} 
\end{enumerate}

We assume that the microscopic system possesses an anti-unitary symmetry involving  time reversal, which we will denote as $\Th$. 
Depending on the system, $\Th$ can be $\CT$, $\CP \CT$, $\CC \CT$, or $\CC\CP\CT$, where $\CT$, $\CP$ and $\CC$ are respectively  time-, parity- and charge-reversal transformations. 
For example, for a system which is only $\CP \CT$ invariant (but not $\CT$, $\CC\CT$,  or $\CC \CP \CT$) we take $\Th=\CP \CT$. 
For a system which is invariant under all of $\CC, \CP$, and  $\CT$,   $\Th$ can be any one of the choices $\CT$, $\CC\CT$, $\CP \CT$ or $\CC\CP\CT$.
Implications of $\Th$ at the macroscopic level are imposed through the dynamical KMS symmetry of~\cite{Glorioso:2017fpd, Glorioso:2016gsa},
which also ensures local thermal equilibrium.  
To leading order in derivatives (or in the classical limit), the transformations act on the two-form fields in the following way\footnote{$\Th \phi (x) \equiv (-1)^\eta \phi (\Th x)$ denotes the transformation of $\phi (x)$ under $\Th$ with $\eta$ the $\Th$-parity of $\phi$.}
\begin{align}\label{DynKMS}
\tilde G_{a\mu\nu} (x) = \Theta G_{a\mu\nu}(x) - i \beta_0 \partial_0 \Theta G_{r\mu\nu} (x) , \quad \tilde G_{r\mu\nu} (x) = \Theta G_{r\mu\nu} (x)  ,
\end{align}
where $\tilde G_{\mu\nu}$ denotes the transformed field.  The transformation~\eqref{DynKMS} is an anti-linear $\mathbb{Z}_2$ transformation. Invariance under it imposes a far-from-equilibrium generalization of the usual KMS conditions and  Onsager's relations on equilibrium correlation functions.

To write down~\eqref{DynKMS} explicitly we need to specify how $G_{\mu\nu}$ transforms under $\CC$, $\CP$ and $\CT$, which should be the same as how the two-form current $J^{\mu\nu}$ transforms under these discrete transformations. 
By definition, $J^{\mu \nu}$ should go to $- J^{\mu \nu}$ under $\CC$. 
 There are two types of parity transformations that we can impose on the $G_{\mu\nu}$ field, which will be respectively referred to as $\CP_+$ and  $\CP_-$ transformations, 
\begin{align}
\CP_+ :&\quad (G_{0i} , G_{ij} ) \to ( - G_{0i} , G_{ij} ) \label{P+} , \\
\CP_- :&\quad (G_{0i} , G_{ij} ) \to ( G_{0i} , - G_{ij} )  \label{P-}   .
\end{align}
Similarly, two types of time-reversal transformations $\CT_+$ and $\CT_-$ act as 
\begin{align}
\CT_+ :&\quad (G_{0i} , G_{ij} ) \to ( G_{0i} , -G_{ij} ) \label{T+} , \\
\CT_- :&\quad (G_{0i} , G_{ij} ) \to ( - G_{0i} ,  G_{ij} )  \label{T-}   .
\end{align}
In the case relevant for electromagnetic plasmas, where the conserved current is \eqref{qed}, the parity and time reversal transformations are $\CP_-$ and $\CT_-$. The alternatives $\CP_+$ and $\CT_+$ could for instance describe the hydrodynamics associated with an alternative theory which had magnetic monopoles but no electric charges, where the conserved current would be $J_{\mu \nu}= F_{\mu \nu}$. See for instance~\cite{monopole_1, monopole_2} for examples of such theories with emergent magnetic monopoles in condensed matter systems.  In principle there could also be theories with a combination of for instance $\CP_-$ and $\CT_+$.

As a result, there are four ways that the two-form current $J^{\mu\nu}$ can transform under combined $\CP_\pm$ and $\CT_\pm$ operations, i.e. under any of $\{\CP_+ \CT_+, \CP_-\CT_+,\CP_+\CT_-,\CP_-\CT_-\}$, depending on its physical interpretation. 
We collect the transformations of $J^{\mu\nu}$, the Stueckelberg field $A_\mu$ and spacetime coordinates in Table \ref{TableDiscrete}.

\begin{table}[ht]
  \caption{Discrete transformations}
\begin{tabular}{ |p{2cm}||p{2cm}|p{3cm}|p{3cm}|p{2cm}|p{2cm}|  }
  \hline
 & $\CC$ & $\CT$ &  $\CP$ & $\CP \CT$ & $\CC \CP \CT$ \\
  \hline
 $x^\mu$  & $x^\mu$  & $(-x^0,  x^i)$  & $(x^0,  -x^i)$  & $- (x^0,  x^i)$    &  $- (x^0,  x^i)$ \\
 $J^{\mu \nu}_{\CP_+ \CT_+} $ & $- J^{\mu \nu}$ & $( J^{0i}, - J^{ij})$ & $(-J^{0i}, J^{ij})$ & $ - J^{\mu \nu}$
 & $J^{\mu  \nu}$  \\
 $J^{\mu \nu}_{\CP_-\CT_+} $ & $- J^{\mu \nu}$ & $( J^{0i}, - J^{ij})$ & $(J^{0i},- J^{ij})$ & $ J^{\mu \nu}$
 & $- J^{\mu  \nu}$  \\
$J^{\mu \nu}_{\CP_+\CT_-} $ & $- J^{\mu \nu}$ & $( - J^{0i}, J^{ij})$ & $(-J^{0i}, J^{ij})$ & $  J^{\mu \nu}$
 & $-J^{\mu  \nu}$  \\ 
 $J^{\mu \nu}_{\CP_-\CT_-} $ & $- J^{\mu \nu}$ & $( -J^{0i}, J^{ij})$  & $(J^{0i}, - J^{ij})$ & $ - J^{\mu \nu}$
 & $J^{\mu  \nu}$  \\
 $A_\mu^{\CP_+ \CT_+} $ &  $- A_\mu$& $(A_0, - A_i)$   & $(A_0, - A_i)$   &  $A_\mu $   &$- A_\mu $\\
 $A_\mu^{\CP_-\CT_+} $ &  $- A_\mu$& $(A_0, - A_i)$   & $(- A_0,  A_i)$   &  $- A_\mu $   &$A_\mu $\\
 $A_\mu^{\CP_+\CT_-} $ &  $- A_\mu$& $(- A_0,  A_i)$  & $(A_0, - A_i)$   &  $- A_\mu $   &$A_\mu$\\
 $A_\mu^{\CP_-\CT_-} $ &  $- A_\mu$& $(-A_0,  A_i)$   & $(-A_0,  A_i)$   &  $A_\mu $   &$- A_\mu $\\
 \hline
\end{tabular}
  \label{TableDiscrete}
\end{table}

With 4 possible choices of $\Theta = \CT, \CP \CT, \CC \CT, \CC\CP\CT$ and 4 possible patterns of discrete $\CP_\pm\CT_\pm$ transformations of $J^{\mu\nu}$ (or $G_{\mu\nu}$), there are potentially many different choices for how to impose the dynamical KMS relations \eqref{DynKMS}. It turns out there are only four inequivalent ones: 
\begin{align} 
\label{11}
\text{KMS}_I:  &\;  (\Th = \CT_-) =  (\Th =\CC \CT_+) , \\
\text{KMS}_{II}: &  \;  (\Th = \CT_+) =  (\Th =\CC \CT_-) , \\
\text{KMS}_{III}: &  \;  (\Th = \CP_+\CT_-) = (\Th =  \CP_- \CT_+) = (\Th = \CC \CP_- \CT_-)  = (\Th = \CC  \CP_+ \CT_+) ,    \\
\text{KMS}_{IV}: &  \;  (\Th = \CP_- \CT_-) = (\Th =  \CP_+ \CT_+) = (\Th = \CC \CP_+ \CT_-)  = (\Th = \CC  \CP_- \CT_+)   .
\label{14}
\end{align}
The corresponding transformations~\eqref{DynKMS} for each of the cases are:
\begin{align} \label{CaseIKMS}
\text{KMS}_I: & &   \tilde{G}_{a 0 i}(\Th x)&=-G_{a 0 i}(x)-i \beta_{0} \partial_{0} G_{r 0 i}(x)  , & \tilde{G}_{r 0 i}(\Th x)&=-G_{r 0 i}(x)   , \nn
&& \tilde{G}_{a i j}(\Th x)&=G_{a i j}(x)+i \beta_{0} \partial_{0} G_{r i j}(x) , & \tilde{G}_{r i j}(\Th x)&=G_{r i j}(x) ,   \\
\label{CaseIIKMS}
\text{KMS}_{II}: & &  \tilde{G}_{a 0 i}(\Th x)&=G_{a 0 i}(x) + i \beta_{0} \partial_{0} G_{r 0 i}(x) , & \tilde{G}_{r 0 i}(\Th x)&=G_{r 0 i}(x), \nn 
&& \tilde{G}_{a i j}(\Th x) &=- G_{a i j}(x)- i \beta_{0} \partial_{0} G_{r i j}(x)  ,   &  \tilde{G}_{r i j}(\Th x)&=- G_{r i j}(x) , \\
\text{KMS}_{III}: & &\tilde{G}_{a\mu \nu} (\Th x) &= G_{a\mu\nu} (x) + i \beta_0 \pt_0 G_{r \mu \nu} (x) , & \tilde{G}_{r\mu\nu} (\Th x) &=  G_{r \mu \nu} (x)  ,\label{CaseIIIKMS} \\
\text{KMS}_{IV}: & & \tilde{G}_{a\mu \nu} (\Th x) &= -G_{a\mu\nu} (x) - i \beta_0 \pt_0 G_{r \mu \nu} (x) , & \tilde{G}_{r\mu\nu} (\Th x) &= - G_{r \mu \nu} (x). \label{CaseIVKMS}
\end{align}

\section{Effective actions}
\label{sec:actions}

Having discussed the field content and symmetries of a hydrodynamic action with a one-form symmetry in Section \ref{sec:FieldsSymm}, we are now ready to write down the most general such CTP action for a dissipative theory of string fluids with an explicitly realized one-form symmetry at non-zero temperature. 
Depending on which combinations of discrete symmetries are preserved by the system at microscopic level, we will find different actions. As mentioned earlier, to impose the dynamical KMS symmetry we assume there is an anti-unitary discrete symmetry $\Th$. 
There can be additional discrete symmetries.
We list all possibilities in Table~\ref{Tab:Dissym}.
 
\begin{table}[!h]
\begin{tabular}{|l|l|l|l|l|l|l|l|}
\hline
                                                       & $\mathcal{C}$ & $\mathcal{P}$ & $\mathcal{T}$ & $\mathcal{C} \mathcal{P}$ & $\mathcal{C} \mathcal{T}$ & $\mathcal{P} \mathcal{T}$ & $\mathcal{C}\mathcal{P}\mathcal{T}$ \\ \hline
\multirow{4}{*}{only $\Theta$}                         &               &               & $\times$      &                           &                           &                           &                                     \\ \cline{2-8} 
                                                       &               &               &               &                           & $\times$                  &                           &                                     \\ \cline{2-8} 
                                                       &               &               &               &                           &                           & $\times$                  &                                     \\ \cline{2-8} 
                                                       &               &               &               &                           &                           &                           & $\times$                            \\ \hline
\multirow{2}{*}{$\Theta$ and $\mathcal{P}$}            &               & $\times$      & $\times$      &                           &                           & $\times$                  &                                     \\ \cline{2-8} 
                                                       &               & $\times$      &               &                           & $\times$                  &                           & $\times$                            \\ \hline
\multirow{2}{*}{$\Theta$ and $\mathcal{C}$}            & $\times$      &               & $\times$      &                           & $\times$                  &                           &                                     \\ \cline{2-8} 
                                                       & $\times$      &               &               &                           &                           & $\times$                  & $\times$                            \\ \hline
\multirow{2}{*}{$\Theta$ and $\mathcal{C}\mathcal{P}$} &               &               & $\times$      & $\times$                  &                           &                           & $\times$                            \\ \cline{2-8} 
                                                       &               &               &               & $\times$                  & $\times$                  & $\times$                  &                                     \\ \hline
$\Theta$, $\mathcal{C}$ and $\mathcal{P}$              & $\times$      & $\times$      & $\times$      & $\times$                  & $\times$                  & $\times$                  & $\times$                            \\ \hline
\end{tabular}
\caption{We show the different possible combinations of discrete symmetries that can appear in the effective action. Each of the 11 rows shows a different allowed combination. For example, fifth row tells us that if we want the only discrete symmetries to be $\Theta$ and $\CP$, then depending on the choice of $\Theta$, there are two possible sets of symmetries that the effective action can have: either $\{ \CP, ~\CT,~ \CP \CT \}$, or  $\{ \CP,~ \CC\CT, ~\CC \CP \CT \}$.}
\label{Tab:Dissym}
\end{table} 
 
When there is more than one conserved anti-unitary symmetry, any of them can be chosen as $\Th$. They lead to equivalent theories and some of the KMS transformations in~\eqref{CaseIKMS}--\eqref{CaseIVKMS} are equivalent. 
For this reason, below we just choose one of these options below. For example, when both $\CP \CT$ and $\CC\CP \CT$ are conserved, we only mention $\Th = \CP \CT$ explicitly, with the understanding that $\Th = \CC\CP \CT$ leads to the same theory. 
The possible hydrodynamic theories can then be classified as follows:

\ben 

\item All discrete symmetries are preserved. In this case we can choose $\Th$ to be any of $\CT$, $\CP \CT$, $\CC \CT$, or $\CC\CP\CT$, which should be all equivalent. From~\eqref{11}--\eqref{14}, all different choices of $\CP_\pm,~ \CT_\pm$ should then yield the same action. There is thus a unique theory. We will denote the corresponding Lagrangian density $\CL_0$. 

\item $\CP$ is conserved but not $\CC$. There are two independent choices of $\Th$,  $\Th = \CT, \CC \CT$. 
Together with possible choices of $\CP_\pm,~ \CT_\pm$, there are four different theories: 
\bega 
(\CP_+ , \Th= \CT_+) =  (\CP_+ , \Th= \CC \CT_-)   , \quad (\CP_+ , \Th= \CT_-)=  (\CP_+ , \Th= \CC \CT_+) , \\
(\CP_- , \Th= \CT_+) =  (\CP_- , \Th= \CC \CT_-)   , \quad (\CP_- , \Th= \CT_-)=  (\CP_- , \Th= \CC \CT_+)   .
\end{gather} 
We will denote the corresponding Lagrangians respectively as $\CL_{\CP_+, \CT_+}$, $\CL_{\CP_+, \CT_-}$, 
$\CL_{\CP_-, \CC \CT_-}$, $\CL_{\CP_-, \CT_-}$. 

\item $\CC$ is conserved but not $\CP$. There are two independent choices of $\Th$,  $\Th = \CT, \CP \CT$, as well as the different choices of $\CP_\pm,~ \CT_\pm$. Due to the various equivalences between different KMS transformations in \eqref{11}--\eqref{14}, we end up with a total of two distinct cases: 
\be
(\CC, \Theta = \CT_- ), \quad (\CC, \Theta = \CP_- \CT_-)  .  
\ee
We call the associated effective actions $\sL_{\CC, \CT_-}$ and  $\sL_{\CC, \CP_-\CT_-}$ respectively.

\item $\CC \CP$ is conserved but not $\CC$ or $\CP$. There are two independent choices of $\Th$,  $\Th = \CT, \CC \CT$, 
as well as the possible choices of $\CP_\pm,~ \CT_\pm$. Since $\CC \CP_\pm$ acts on $G_{\mu \nu}$ as $\CP_\mp$ we get the same four theories as in $\CP$ conserved case, i.e.,\footnote{Here we have underlined the notation we will be using from now on for these effective actions, in two cases replacing the notation introduced in point 2. Since the $\CP_-$, $\CT_-$ symmetries apply to the case of an electromagnetic plasma and therefore have the best-understood physical interpretation, whenever we have an option between labelling an action with the $\CP_-, \CT_-$ symmetries and labelling it with $\CP_+$, $\CT_+$ or some combination such as $\CP_+$, $\CT_-$, we will choose to label it with  $\CP_-, \CT_-$. In fact, we will find that all possible effective actions can be labelled with combinations involving $C$, $\CP_-$, and $\CT_-$.} 
\begin{align}
& \sL_{\CC \CP_+, \CT_+} = \underline{\bf \sL_{\CP_-, \CC \CT_-}} ,  \\
& \underline{\bf \sL_{\CC \CP_-, \CC \CT_-}} = \sL_{\CP_+, \CT_+} ,\\
&  \sL_{\CC \CP_+, \CT_-} = \underline{\bf \sL_{\CP_- , \CT_-}} ,\\
& \underline{\bf \sL_{\CC \CP_-, \CT_-}} = \sL_{\CP_+,\CT_- }.
\end{align} 

\item $\Th$ is the only discrete symmetry. In this case there are altogether four different theories corresponding to the four possible KMS 
transformations~\eqref{11}--\eqref{14}. 
We will denote the corresponding Lagrangians respectively as $\CL_{\CC\CT_-}$, $\CL_{\CT_-}$, 
$\CL_{\CC \CP_-  \CT_-}$, $\CL_{ \CP_- \CT_-}$.

\een

To summarize, based on the above classification, we have ended up with eleven different possible effective actions. We can choose to represent each of these by specifying which discrete symmetries are satisfied assuming that the parity and time reversal transformations are $\CP_-$ and $\CT_-$: 
\begin{align} \label{list_1}
& \sL_0 = \sL_{\CC, \CT_-, \CP_-}, \\
& \CL_{\CT_-}, \quad \CL_{\CC\CT_-}, \quad  \CL_{\CC \CP_- \CT_-}, \quad \CL_{\CP_- \CT_-}, \\
& \CL_{\CP_-, \CC \CT_-}, \quad \CL_{\CP_-, \CT_-}, \\ 
& \CL_{\CC, \CT_-}, \quad  \CL_{\CC, \CP_- \CT_-} ,\\
&  \CL_{\CC \CP_-, \CT_-}, \quad  \CL_{\CC \CP_-, \CC \CT_-} . \label{list_2}
\end{align} 

When parity or time reversal acts as $\CP_+$ or $\CT_+$, the actions for each of the different choices of discrete symmetry can be mapped to one of the cases in \eqref{list_1}--\eqref{list_2}, and the mapping can be inferred using  
 points 1 through 5 above, together with \eqref{11}--\eqref{14}. 
 
Below, we will write the effective actions for each of these eleven cases to all orders in the field $G_{r0i}$, and perform a systematic expansion in derivatives.  We will see that all theories are dominated by diffusion at leading order, so that in the derivative counting we should assign weight $2$ to $\p_0$ and  weight $1$ to $\p_i$. Since  $G_{r0i}$ is assigned weight 0, by \eqref{FieldRel1}, $G_{rij}$ has weight 1. Then since both sides of the dynamical KMS relations \eqref{DynKMS} should have the same weight,  $G_{a0i}$ and $G_{aij}$ have weights $2$ and  $1$ respectively. 
 
 We will write the actions in each of the symmetry cases up to weight 3. We state the results here, leaving details to Appendix~\ref{App:KMS}.

 For each of the symmetry cases, we find that the action starts with terms of weight 2. It turns out that for each of the eleven cases in \eqref{list_1}--\eqref{list_2}, the weight 2 part of the action, $\sL^{(2)}$, is always one out of two options.  Since the weight two part of the action will turn out to determine the nature of the diffusive transport in the system, we will label the two options $\sL^{(2)}_{\rm D1}$ and $\sL^{(2)}_{\rm D2}$ respectively.   In all cases with any discrete symmetry involving charge conjugation the weight 2 part of the effective action is $\sL^{(2)}_{\rm D1}$, while in all cases without any symmetry involving charge conjugation, we have $\sL^{(2)}_{\rm D2}$. That is, 
 \begin{align} 
 &\sL_{\rm D1}^{(2)} = \sL_{\CC, \CP_-, \CT_-}^{(2)}=\sL_{\CP_-, \CC \CT_-}^{(2)}  =  \sL_{\CC\CP_-, \CC \CT_-}^{(2)}  =  \sL_{\CC\CP_-,  \CT_-}^{(2)}  = \sL_{\CC,  \CT_-}^{(2)}  = \sL_{\CC, \CP_- \CT_-}^{(2)}  =  \sL_{\CC \CT_-}^{(2)}  = \sL_{\CC \CP_- \CT_-}^{(2)} , \label{d1cases} \\
 &\sL_{\rm D2}^{(2)} = \sL_{\CP_-, \CT_-}^{(2)}  =  \sL_{ \CT_-}^{(2)}  =  \sL_{\CP_- \CT_-}^{(2)} .  \label{d2cases}
    \end{align} 
Note in particular that for describing magnetohydrodynamics in neutron stars, we should treat the charge conjugation symmetry as broken, as the energy is high enough that we can treat electrons as relativistic, but not high enough for the existence of positrons. Depending on whether we assume that the effect of parity breaking due to weak interactions is significant, the relevant effective action may be $\sL_{\CP_-, \CT_-}$ or  $\sL_{ \CT_-}$, both of which have the same leading diffusive transport behaviours due to \eqref{d2cases}.

 $\sL^{(2)}_{\rm D1}$ and  $\sL^{(2)}_{\rm D2}$ are given as follows (Note that in these and all other equations below, coefficients denoted by small letters such as $a, d, \tilde d$ are arbitrary real functions of $G_{r0i}^2$ unless specified otherwise.): 
 \begin{align} 
\CL_{\rm D1}^{(2)} &= a \,  G_{r0i} G_{a0i}   + (d \, \delta_{ik} \delta_{jl}  + \tilde d \, \epsilon_{ijm} {\epsilon}_{kln} G_{r0m} G_{r0n}) \,  G_{aij} \,  ( i G_{akl} - \beta_0  \pt_0 G_{rkl} ), \label{L2_d1}
 \end{align} 
and
  \begin{align} 
\CL_{\rm D2}^{(2)} &= \CL_{\rm D1}^{(2)} + p  \delta_{ik} \epsilon_{jln} G_{r0n} G_{aij} \partial_0 G_{rkl}  .\label{L2_d2}
 \end{align} 
 
 Using \eqref{jmn}, the constitutive relations from these actions are respectively 
 \begin{align}\label{j10}
J_{\rm D1}^{0i} &= a \,  G_{r0i}  , \\
J_{\rm D1}^{ij} &= 
        -2 \beta_0 (d \, \delta_{ik} \delta_{jl}  + \tilde d \, \epsilon_{ijm} {\epsilon}_{kln} G_{r0m} G_{r0n}) \pt_0 G_{rkl} , \label{j1ij}
\end{align}
and 
\be 
J_{\rm D2}^{0i} = J_{\rm D1}^{0i}, \quad  J_{\rm D2}^{ij}  = J_{\rm D1}^{ij}  + p  (\delta_{ik} \epsilon_{jln} - \delta_{jk} \epsilon_{iln}) G_{r0n} \partial_0 G_{rkl}  .  \label{j2}
 \ee
Note that in the discussion of this paper, we set the fields $G_{a\mu \nu}$ to zero in the constitutive relations, which corresponds to neglecting the effects of statistical or quantum noise. 
We will discuss the physical consequences of these constitutive relations for the diffusive transport behaviour of the system in Section \ref{sec:EFT-Sym}. Readers who are only interested in those aspects can directly jump to that section.

We now separately list the effective actions up to weight 3 for the different symmetry cases. In the expressions below, tensors such as $F_{ijk}$, $O_{ijk}$, $V_{ijk}$ are all constructed from  combinations of $G_{r0i}$, $\delta_{ij}$, and $\epsilon_{ijk}$ and have weight zero. The $(o)$ and $(e)$ superscripts specify whether the tensor contains an odd or even total number of factors of $G_{r0i}$. In cases where no superscripts are specified, we can have arbitrary numbers of factors of $G_{r0i}$ in the tensor.   We can have arbitrary combinations consistent with these conditions and the tensor structures, except in some special cases where we specify that the tensors are constants with respect to $G_{r0i}$ or specify their symmetries under exchange of indices. In Sec. \ref{sec:weight_three}, we will discuss more explicitly which of the weight 3 terms are relevant for the linear response theory and can lead to cubic corrections to the dispersion relations.

\subsection{Invariant under all discrete symmetries}

We first consider the simplest case: the underlying system (i.e.,~the microscopic Lagrangian) is invariant separately under all  discrete spacetime symmetries. In this case, all different  choices of $\CP_\pm,~ \CT_\pm$ and $\Th$ yield the same action at least up to weight 3, which we call $\sL_0$. The weight 3 part of the action turns out to be zero imposing the KMS conditions. We have 
\begin{align}\label{minac0}
\CL_0^{(2)} &=  \sL^{(2)}_{\rm D1} , \quad \CL_0^{(3)} =  0 , 
\end{align}
so that the constitutive relation for this case is given by \eqref{j10}--\eqref{j1ij}, with no corrections from weight 3 terms. 

\subsection{No discrete symmetry other than $\Th$}
\label{sec:nodisc}

Now  consider the other extreme: no discrete symmetry is present other than $\Th$. 
In this case, there are four possible theories corresponding to four different ways~\eqref{11}--\eqref{14} of imposing the dynamical KMS symmetry. They are given as follows. 
\subsubsection{ $(\Th = \CT_-) =  (\Th =\CC \CT_+)$} 

In this case we use KMS$_I$ as given in~\eqref{CaseIKMS}. The resulting Lagrangian is:  
\begin{align} 
  &\CL_{\CT_-}^{(2)}  =  \sL_{\rm D2}^{(2)}  ,
\label{oo1} 
\end{align} 
and
\begin{align} 
 \CL_{\CT_-}^{(3)}  = ~&F_{ijk}^{(e)} G_{a0i} \partial_j G_{r0k}  \nn 
& + O^{(o)}_{ijk} ( - i \frac{2}{\beta_0} G_{a0i} G_{ajk} +    G_{ajk} \partial_0 G_{r0i}  + G_{a0i} \partial_0 G_{rjk} ) \nn 
&+  O^{(e)}_{ijk} ( G_{a0i} \partial_0 G_{rjk} - G_{ajk} \partial_0 G_{r0i} ) 
\nn
&+ V_{ijklm}^{(e)} ( - i \frac{2}{\beta_0}G_{aij} \partial_k G_{alm}  +   G_{aij} \partial_k \partial_0 G_{rlm} + \partial_0 G_{rij} \partial_k G_{alm} ) \nn 
& + V^{(o)}_{ijklm} (G_{aij} \partial_k \partial_0 G_{rlm}  -  \partial_0 G_{rij} \partial_k G_{alm} )  
 \nn 
 & + C^{(o)}_{ijklmn} (G_{aij}  G_{akl} \partial_{(m} G_{r0n)} + i \beta_0   
 G_{aij} \partial_0 G_{rkl} \partial_{(m} G_{r0n)}) \nn 
& + Q_{ijklmn}^{(e)}G_{aij} \partial_0 G_{rkl} \partial_{(m} G_{r0n)} 
  \nn 
 & +  R_{ijklmn}^{(e)} (G_{aij} G_{akl} G_{amn} + \frac{3}{2} i \beta_0 G_{aij} G_{akl} \partial_0 G_{rmn} - \frac{\beta_0^2}{2}  G_{aij} \partial_0 G_{rkl} \partial_0 G_{rmn}) \nn 
 & + S^{(o)}_{ijklmn} (G_{aij} G_{akl} \partial_0 G_{rmn} +  i \beta_0 G_{aij} \partial_0 G_{rkl} \partial_0 G_{rmn}) .
\end{align} 
Here, $F^{(e)}$ is a constant independent of $G_{r0i}^2$, and $F^{(e)}_{ijk}= - F^{(e)}_{kji}$. $Q^{(e)}$ is the most general tensor structure consistent with the condition $Q^{(e)}_{ijklmn}=-Q^{(e)}_{klijmn}$.

\subsubsection{ $ (\Th =\CC \CT_-) = (\Th = \CT_+) $} 

In this case we use KMS$_{II}$ as given in~\eqref{CaseIIKMS}. The resulting Lagrangian is:  
\be 
  \CL_{\CC\CT_-}^{(2)}  = \sL^{(2)}_{\rm D1} ,
\ee
and
\begin{align} 
  \CL_{\CC\CT_-}^{(3)}  = ~&  F_{ijk} G_{a0i} \partial_j G_{r0k}  \nn 
&  +  O_{ijk} (G_{a0i} \partial_0 G_{rjk}- G_{ajk} \partial_0 G_{r0i}  ) 
\nn
&+ V_{ijklm} ( - i \frac{2}{\beta_0} G_{aij} \partial_k G_{alm}  +  G_{aij} \partial_k \partial_0 G_{rlm} +  \partial_0 G_{rij} \partial_k G_{alm} ) 
 \nn 
 & + C_{ijklmn} (G_{aij}  G_{akl} \partial_{(m} G_{r0n)} + i \beta_0   
  G_{aij} \partial_0 G_{rkl} \partial_{(m} G_{r0n)}) \nn 
   & +  R_{ijklmn} (G_{aij} G_{akl} G_{amn} + \frac{3}{2} i \beta_0 G_{aij} G_{akl} \partial_0 G_{rmn} - \frac{\beta_0^2}{2}  G_{aij} \partial_0 G_{rkl} \partial_0 G_{rmn}) ,
\end{align} 
where $F$ is a constant independent of $G_{r0i}^2$, and $F_{ijk}= - F_{kji}$.

\subsubsection{$(\Th = \CC \CP_- \CT_-) = (\Th = \CP_+\CT_-) = (\Th =  \CP_- \CT_+)   = (\Th = \CC  \CP_+ \CT_+)$}

In this case we use KMS$_{III}$ as given in~\eqref{CaseIIIKMS}. The resulting Lagrangian is:  
\be
  \CL_{\CC \CP_- \CT_-}^{(2)}  =   \CL_{\rm D1}^{(2)},
  \ee
and  
\begin{align} 
   \CL_{\CC \CP_- \CT_-}^{(3)} =  ~ &   O_{ijk} ( - i \frac{2}{\beta_0}G_{a0i} G_{ajk} +    G_{ajk} \partial_0 G_{r0i}  + G_{a0i} \partial_0 G_{rjk} )
\nn
& + V_{ijklm} (G_{aij} \partial_k \partial_0 G_{rlm}  -  \partial_0 G_{rij} \partial_k G_{alm} )  
 \nn 
& + Q_{ijklmn}G_{aij} \partial_0 G_{rkl} \partial_{(m} G_{r0n)} 
  \nn 
 & +  R_{ijklmn} (G_{aij} G_{akl} G_{amn} + \frac{3}{2} i \beta_0 G_{aij} G_{akl} \partial_0 G_{rmn} - \frac{\beta_0^2}{2}  G_{aij} \partial_0 G_{rkl} \partial_0 G_{rmn}) , \label{cpmtm1}
 \end{align} 
where $Q_{ijklmn} = - Q_{klijmn}$.

\subsubsection{$(\Th = \CP_- \CT_-) = (\Th =  \CP_+ \CT_+) = (\Th = \CC \CP_+ \CT_-)  = (\Th = \CC  \CP_- \CT_+) $}   

In this case we use KMS$_{IV}$ as given in~\eqref{CaseIVKMS}. The resulting Lagrangian is:  
\be 
\CL_{\CP_- \CT_-}^{(2)}  =  \CL_{\rm D2}^{(2)},  
\ee
and
\begin{align} 
     \CL_{\CP_- \CT_-}^{(3)}  = &F_{ijk}^{(o)} G_{a0i} \partial_j G_{r0k}  \nn 
& + O^{(e)}_{ijk} ( - i \frac{2}{\beta_0}G_{a0i} G_{ajk} +    G_{ajk} \partial_0 G_{r0i}  + G_{a0i} \partial_0 G_{rjk} ) \nn 
& +  O^{(o)}_{ijk} ( G_{a0i} \partial_0 G_{rjk}- G_{ajk} \partial_0 G_{r0i} ) 
\nn
&+ V_{ijklm}^{(o)} (- i \frac{2}{\beta_0} G_{aij} \partial_k G_{alm}  +  G_{aij} \partial_k \partial_0 G_{rlm} +  \partial_0 G_{rij} \partial_k G_{alm} ) \nn 
& + V^{(e)}_{ijklm} (G_{aij} \partial_k \partial_0 G_{rlm}  -  \partial_0 G_{rij} \partial_k G_{alm} )  
 \nn 
 & + C^{(e)}_{ijklmn} (G_{aij}  G_{akl} \partial_{(m} G_{r0n)} + i \beta_0   
 G_{aij} \partial_0 G_{rkl} \partial_{(m} G_{r0n)}) \nn 
& + Q_{ijklmn}^{(o)}G_{aij} \partial_0 G_{rkl} \partial_{(m} G_{r0n)} 
  \nn 
 & +  R_{ijklmn}^{(o)} (G_{aij} G_{akl} G_{amn} + \frac{3}{2} i \beta_0 G_{aij} G_{akl} \partial_0 G_{rmn} - \frac{\beta_0^2}{2}  G_{aij} \partial_0 G_{rkl} \partial_0 G_{rmn}) \nn 
 & + S^{(e)}_{ijklmn} (G_{aij} G_{akl} \partial_0 G_{rmn} +  i \beta_0 G_{aij} \partial_0 G_{rkl} \partial_0 G_{rmn}) ,
\label{oo4}
\end{align} 
where $F^{(o)}$ is a constant independent of $G_{r0i}$, $F^{(o)}_{ijk} = - F^{(o)}_{kji}$, and $Q_{ijklmn}^{(o)}=-Q_{klijmn}^{(o)}$.

\subsection{$\CP$ is conserved, $\CC$ is not conserved} 
\label{pnotc}

The actions $\sL_{\CP_-, \CT_-}$ and $\sL_{\CP_-, \CC\CT_-}$ in this case can be obtained from $\sL_{ \CT_-}$ and $\sL_{\CC_-\CP_-\CT_-, \CC\CT_-}$ respectively of the previous section by imposing parity conservation.

\subsubsection{$(\CP_- , \Th= \CT_-)$} 

In this case, the Lagrangian can be obtained from \eqref{oo1} by imposing $\CP_-$ invariance 
which gives
\begin{align}
\CL_{\CP_- , \CT_-}^{(2)} &= \CL^{(2)}_{\rm D2}, \quad  \CL_{\CP_- , \CT_-}^{(3)}=0 . 
\end{align}

\subsubsection{$(\CP_- , \Th= \CC\CT_-)$} 

In this case the Lagrangian can be obtained from~\eqref{cpmtm1} by imposing $\CP_-$ invariance. The result 
turns out to be the same as~\eqref{minac0}, i.e. $\CL_{\CP_-, \CC \CT_-} = \CL_0$.

\subsection{$\CC$ conserved, $\CP$ is not conserved}
\label{cnotp}

The actions $\CL_{\CC, \CT_-}$ and  $\CL_{\CC, \CP_-\CT_-}$  in this case can be obtained by from~$\sL_{\CC \CT_-}$ and $\sL_{\CC_- \CP_-\CT_-}$  by imposing $\CC$ conservation. 

\subsubsection{$(\CC, \Th= \CT_-)$ } 

This action can be obtained by imposing $\CC$ invariance in~\eqref{oo1}. We find
\be
\CL_{\CC, \CT_-}^{(2)} =\CL_{\rm D1}^{(2)}  ,
\ee
and
\begin{align}  
 \CL_{\CC, \CT_-}^{(3)} = ~&  F^{(e)}_{ijk} G_{a0i} \partial_j G_{r0k}  \nn 
&  +  O^{(e)}_{ijk} (G_{a0i} \partial_0 G_{rjk}- G_{ajk} \partial_0 G_{r0i}  ) 
\nn
&+ V^{(e)}_{ijklm} ( - i \frac{2}{\beta_0} G_{aij} \partial_k G_{alm}  +  G_{aij} \partial_k \partial_0 G_{rlm} +  \partial_0 G_{rij} \partial_k G_{alm} ) 
 \nn 
 & + C^{(o)}_{ijklmn} (G_{aij}  G_{akl} \partial_{(m} G_{r0n)} + i \beta_0   
  G_{aij} \partial_0 G_{rkl} \partial_{(m} G_{r0n)}) \nn 
   & +  R^{(o)}_{ijklmn} (G_{aij} G_{akl} G_{amn} + \frac{3}{2} i \beta_0 G_{aij} G_{akl} \partial_0 G_{rmn} - \frac{\beta_0^2}{2}  G_{aij} \partial_0 G_{rkl} \partial_0 G_{rmn}) , 
\end{align} 
where $F^{(e)}$ is a constant independent of $G_{r0i}$, and $F^{(e)}_{ijk} = - F^{(e)}_{kji}$.

\subsubsection{$(\CC, \Th= \CP_-\CT_-)$ }
This case can be obtained  by imposing $\CC$ invariance in~$\sL_{\CC \CP_- \CT_- }$. We find  
\be
\CL_{\CC, \CP_-\CT_-}^{(2)} =\CL_{\rm D1}^{(2)} , 
\ee 
and
 \begin{align} 
   \CL_{\CC, \CP_- \CT_-}^{(3)} =  ~ &   O^{(e)}_{ijk} ( - i \frac{2}{\beta_0}G_{a0i} G_{ajk} +    G_{ajk} \partial_0 G_{r0i}  + G_{a0i} \partial_0 G_{rjk} )
\nn
& + V^{(e)}_{ijklm} (G_{aij} \partial_k \partial_0 G_{rlm}  -  \partial_0 G_{rij} \partial_k G_{alm} )  
 \nn 
& + Q^{(o)}_{ijklmn}G_{aij} \partial_0 G_{rkl} \partial_{(m} G_{r0n)} 
  \nn 
 & +  R^{(o)}_{ijklmn} (G_{aij} G_{akl} G_{amn} + \frac{3}{2} i \beta_0 G_{aij} G_{akl} \partial_0 G_{rmn} - \frac{\beta_0^2}{2}  G_{aij} \partial_0 G_{rkl} \partial_0 G_{rmn}) ,   \label{cpmtm}
 \end{align} 
where $Q_{ijklmn}^{(o)}=-Q_{klijmn}^{(o)}$. 

\subsection{With $\CC \CP$ but neither $\CC$ nor $\CP$}
\subsubsection{$(\CC\CP_- , \Th= \CT_-)$} 

 In this case the Lagrangian can be obtained from~$\sL_{C \CP_- \CT_-}$ by imposing $\CC\CP_-$ invariance. We find   
 \be 
 \CL_{\CC \CP_- , \CT_-}^{(2)}= \sL^{(2)}_{\rm D1},
 \ee
and
\begin{align}
\CL_{\CC \CP_- , \CT_-}^{(3)}=~&   O^{(o)}_{ijk} ( - i \frac{2}{\beta_0}G_{a0i} G_{ajk} +    G_{ajk} \partial_0 G_{r0i}  + G_{a0i} \partial_0 G_{rjk} )
\nn
& + V^{(o)}_{ijklm} (G_{aij} \partial_k \partial_0 G_{rlm}  -  \partial_0 G_{rij} \partial_k G_{alm} )  
 \nn 
& + Q^{(e)}_{ijklmn}G_{aij} \partial_0 G_{rkl} \partial_{(m} G_{r0n)} 
  \nn 
 & +  R^{(e)}_{ijklmn} (G_{aij} G_{akl} G_{amn} + \frac{3}{2} i \beta_0 G_{aij} G_{akl} \partial_0 G_{rmn} - \frac{\beta_0^2}{2}  G_{aij} \partial_0 G_{rkl} \partial_0 G_{rmn}),  \label{cpmtm}
\end{align}
where $Q_{ijklmn}^{(e)}=-Q_{klijmn}^{(e)}$.

\subsubsection{$(\CC\CP_- , \Th= \CC\CT_-)$} 

In this case, the effective Lagrangian can be obtained from $\sL_{\CP_- \CT_-}$ by imposing $\CC\CP_-$ invariance. We get
\begin{align}
\CL_{\CC \CP_- ,\CC \CT_-}^{(2)} = \CL_{\rm D1}^{(2)},
\end{align}
and 
\begin{align}
\CL_{\CC \CP_- ,\CC \CT_-}^{(3)} =~ &F_{ijk}^{(o)} G_{a0i} \partial_j G_{r0k}  \nn 
& +  O^{(o)}_{ijk} ( G_{a0i} \partial_0 G_{rjk}- G_{ajk} \partial_0 G_{r0i} ) 
\nn
&+ V_{ijklm}^{(o)} (- i \frac{2}{\beta_0} G_{aij} \partial_k G_{alm}  +  G_{aij} \partial_k \partial_0 G_{rlm} +  \partial_0 G_{rij} \partial_k G_{alm} ) \nn 
 & + C^{(e)}_{ijklmn} (G_{aij}  G_{akl} \partial_{(m} G_{r0n)} + i \beta_0   
 G_{aij} \partial_0 G_{rkl} \partial_{(m} G_{r0n)}) \nn 
 & + S^{(e)}_{ijklmn} (G_{aij} G_{akl} \partial_0 G_{rmn} +  i \beta_0 G_{aij} \partial_0 G_{rkl} \partial_0 G_{rmn}) ,
\label{oo4}
\end{align}
where $F^{(o)}$ is a constant independent of $G_{r0i}$, and $F^{(o)}_{ijk} = - F^{(o)}_{kji}$.

\section{Diffusion behaviours of a one-form charge density}\label{sec:EFT-Sym}

We now proceed to study the characteristics of transport phenomena in the theories constructed in the previous section. 

For the purpose of analyzing transport behaviors at the level of linear response, 
we assume that $G_{r0i}$ has a background value $\mu_i$ which is constant with respect to $\vx$ and $t$, and that it can have small fluctuations $f_{r0i}$ around this value, 
\be 
G_{r0i} = \mu_i + f_{r0i}  . \label{mudef} 
\ee
$f_{r0i}$ can be seen as including both fluctuations in the source fields $b_{r0i}$ and the contribution $\partial_0 A_i - \partial_i A_0$ from the dynamical fields. In much of the discussion below, we will set the fluctuations in $b_{r0i}$  to zero, so that 
\be 
f_{r0i} = \partial_0 A_i - \partial_i A_0 
\ee
and 
\be 
\partial_0 G_{rij} = H_{0ij} + \partial_i G_{r0j} - \partial_j G_{r0i} =\partial_i f_{r0j} - \partial_j f_{r0i} . 
\ee
We can then expand all coefficients appearing in the effective actions (which are functions of $G_{r0i}^2$) around their values at $\mu^2$, in order to obtain expressions for $J^{\mu \nu}$ up to linear order in the fluctuations $f_{r0i}$.

For convenience we take the equilibrium vector chemical potential $\mu_i$ to point along the $z$-axis with the magnitude $\mu$ and use Latin indices from the beginning of the alphabet $a,b,\ldots$ to denote the remaining two spatial directions, $x$ and $y$. We can decompose the current into $J^{\mu\nu}$ into $J^{0i} = (J^{0z},J^{0a})$ and  $J^{ij} = (J^{za}, J^{ab})$, where we write $J^{ab} = \ep_{ab} \hat J^z$ with $\ep_{ab} =  \ep_{zab}$, and  $\hat J^z = \half \ep_{ab} J^{ab}$. A similar decomposition can be done for the background field $b_{ij}$, into $b_{r0i} = (b_z, b_a)$, $E_{ij} = - H_{r0ij} = (E_{za}, \hat E_z)$, and $B_{zab} = H_{zab}$. $E_{ij}$ and $B_{zab}$ can be seen as higher-form analogs of external electric and magnetic fields, and should not be confused with the physical electric and magnetic fields in MHD. To keep the discussion of this section general and applicable to any string fluid, we will not assume any particular physical interpretation of $J^{\mu \nu}$, and for the most part will not explicitly refer to physical electric and magnetic fields in MHD.

Three key transport coefficients that appear in the discussion below are the higher-form ``conductivities'' $\sigma^{\para}$, $\sigma^{\perp}_1$ and $\sigma^{\perp}_2$. The linear response interpretation of these and other coefficients in terms of retarded Green's functions is discussed in detail in Appendix C of \cite{Vardhan:2022wxz} (see in particular equation (C24)). By comparing these Green's functions with those that appear in the discussion of string fluids in Sec.~5.4 of \cite{Armas:2018zbe} derived using an alternative approach, we find that $\sigma^{\para}$, $\sigma^{\perp}_1$ and $\sigma^{\perp}_2$ can be identified respectively as the coefficients $r_{\para}$, $r_{\perp}$ and $\tilde r_{\perp}$ appearing in that reference.

\subsection{Equations of motion in D1 case} \label{sec:cpcon}

\subsubsection{Diffusion equations}

In cases where the effective action up to weight 2 is given by $\sL^{(2)}_{\rm D1}$, the constitutive relation is \eqref{j10}--\eqref{j1ij}. Using \eqref{mudef} and expanding various coefficients in the effective action to get $J^{\mu \nu}$ up to linear order in $f_{r0i}$, we find  
\begin{align}
J_{\rm D1}^{0i} &= a \mu_i +  (a \delta_{ij} + \tilde a \mu_i \mu_j  ) f_{r0j}, \label{j1} \\
 J_{\rm D1}^{ij} &= 
        -2 \beta_0 (d \delta_{ik} \delta_{jl} 
    + \tilde d\, \Tilde{\epsilon}_{ij} \Tilde{\epsilon}_{kl} ) \pt_0 G_{rkl}, \label{JC-4-2}
\end{align}
where $\tilde{a}= 2 \frac{d a(G_{r0i}^2)}{d(G_{r0i}^2)}\mid_{\mu^2}$, $\tilde{\epsilon}_{ij} \equiv \epsilon_{ijk} \mu_k$, and the coefficients in \eqref{JC-4-2} should now be seen as constants evaluated at $G_{r0i}^2= \mu^2$. Recall from \eqref{d1cases} that if $\CP, \CT$ are taken to be $\CP_-, \CT_-$ in \eqref{P-def},\eqref{T-def}, then all cases with any symmetry involving $\CC$ lead to these constitutive relations at leading order in derivatives.

For the rest of this subsection, to simplify the notation we will drop the D1 subscript in the currents. Now consider the physical interpretation of various coefficients in~\eqref{j1}--\eqref{JC-4-2}.
The static susceptibilities can be defined as 
\be 
J^{0z} = \chi^{\para} b_z, \quad J^{0a} = \chi^\perp_{ab} b_b  .
\ee
From~\eqref{j1}  we have
\begin{align}
\chi^{\parallel} = a + \tilde a \mu^2 , \quad \chi^{\perp}_{ab} \equiv \chi_\perp \delta_{ab} = a \de_{ab}  . 
\label{susceptibilities}
\end{align}
$\chi^\perp$ must be diagonal as an off-diagonal term would be incompatible with the three-dimensional rotational symmetry of the underlying theory. We see that coefficient $\tilde a$ controls the asymmetry between the directions longitudinal and transverse to the direction of $\mu^i$.  The ``conductivities'' under an external ``electric'' field  can be defined as 
\be 
J^{za} = \sig^\perp_{ab} E_{zb}, \quad \hat J^z = \sig^\para \hat E_z   .
\ee
From~\eqref{JC-4-2}, we find that 
\begin{align}
\sigma^{\perp}_{ab} \equiv \sigma_1^\perp \delta_{ab}=  2 \beta_0 d  \de_{ab}, \quad \sigma^{\para} = 2\beta_0(d+2 \tilde d\mu^2) ,
\label{conductivities}
\end{align}
with $\tilde d$ controlling the asymmetry between conductivities in the directions parallel and perpendicular to $\mu^i$. 
One can check that from the condition \eqref{cons3}, $\sigma^{\parallel}$ and $\sigma^{\perp}_1$ are both non-negative.

The equations of motion are simply the conservation equations 
\begin{align} \label{Eom1}
\partial_j J^{j 0} &= 0  ,   \\ 
\partial_0 J^{0i}+ \partial_{j} J^{ji}&=0  .
\label{Eom2}
\end{align}
Let us first consider $\mu =0$ for which 
\begin{align}
J^{0i} &=  a f_{r0i} , \label{Mu-0-J0i}\\
J^{ij} &=  -2 \beta_0 d \left( \partial_i f_{r0j} - \partial_j f_{r0i}\right) = - {2 \beta_0 d \ov a} (\p_i J^{0j} - \p_j J^{0i}) + \cdots, \label{Mu-0-Jij}
\end{align}
where in the second equality we have only kept the lowest order term in derivatives. Equation~\eqref{Eom1} then reduces to $\p_j f_{r0j} = 0$, while~\eqref{Eom2} leads to the standard diffusion equation 
\begin{align}\label{EqDiff0}
\partial_0 J^{0i} - D \partial^2_j J^{0i} = 0 ,  \quad D =  {2 \beta_0 d \ov a} = \frac{\sigma}{\chi}, \quad \sig = 2 \beta_0 d , \quad \chi =a   .
\end{align}
With $\mu =0$,  there is only one conductivity $\sig = \sigma_1^\perp$ and susceptibility $\chi = \chi_{\perp}$. 

Now consider $\mu \neq 0$. At leading order in derivatives we have 
\begin{align}
J^{0z} =a \mu +  \chi_\parallel f_{r0z} , \quad J^{0a} = \chi_\perp f_{r0a},
\end{align}
which are related by Eq.~\eqref{Eom1}, i.e. $\p_z J^{0z} + \p_a J^{0a} =0 $. The spatial components of the current can then be written as  
\begin{align}
J^{za} &= - \sig_1^\perp (\p_z f_{r0a} - \p_a f_{r0z}) = - \sig_1^\perp \le({\p_z J^{0a} \ov  \chi_\perp} - {\p_a J^{0z} \ov \chi_\parallel} \ri), \\ 
\hat J^z &= \half \ep_{ab} J^{ab} =- \sig^\para  \ep_{ab} \p_a f_{r0b} = - {\sig^\para  \ov  \chi_\perp }  \ep_{ab} \p_a J^{0b}. 
\end{align} 
Eq.~\eqref{Eom2} for $i=z$  then leads to 
\begin{align} \label{d0}
\p_0 J^{0z} - \le( D_\para \p_z^2 + D_\perp  \p_a^2 \ri) J^{0z} = 0, \quad D_\para = {\sig_1^{\perp}   \ov \chi_\perp}, \quad D_\perp = {\sig_1^{\perp} \ov  \chi_\para}  ,
\end{align}
which leads to dispersion relation 
\be 
\om = - i (D_{\para} k_z^2 + D_\perp k_a^2) \ . \label{disp1}
\ee
Eq.~\eqref{Eom2} for $i=a$  gives 
\begin{align} \label{d1}
\p_0 J^{0a} - {1 \ov \chi_\perp} \le[ \sig_1^\perp  \p_z^2  + \sig^\para \p_b^2 \ri] J^{0a} + 
\le({\sig^\para \ov \chi_\perp} - {\sig_1^\perp \ov \chi_\para} \ri) \p_a \p_b J^{0b} = 0  .
\end{align}
Acting with $\p_a$ on the above equation, we find a diffusive equation for $\p_a J^{0a} $ which is the same as~\eqref{d0}. This can be expected from~\eqref{Eom1} as  $\p_a J^{0a} = - \p_z  J^{0z}$.  Projecting~\eqref{d1} to 
$J^{0a}_\perp$, the component perpendicular to $k^a$, we find a diffusion equation
\be \label{d2} 
\le(\p_0 - D_\para \p_z^2 - \tilde D_\perp \p_b^2 \ri)  J^{0a}_\perp  = 0, \quad \tilde D_\perp = {\sig^\para  \ov \chi_\perp} ,
\ee
with 
dispersion relation 
\be 
\om = - i (D_{\para} k_z^2 + \tilde D_\perp k_a^2) . \label{disp2}
\ee

To summarize, due to the constraint~\eqref{Eom1},  we have two independent decoupled components of $J^{0i}$: one is proportional to $J^{0a}_\perp$ and satisfies diffusion equation~\eqref{d2}; the other is a combination of $J^{0z}$ and the longitudinal component of $J^{0a}$ (i.e. the part proportional to $\p_a J^{0a}$), and satisfies the diffusion equation~\eqref{d0}. More explicitly, 
we can write these two components in  momentum space as
\bega 
\hat J^\al  (k^\mu) = \bfe^{\al}_i J^{0i} (k^\mu), \quad \al =1, 2, \quad \bfe_i^\al k^i = 0, \quad
\bfe^\al \cdot \bfe^\beta = \de^{\al \beta} , \quad k_\perp =  \sqrt{k_x^2 + k_y^2} ,\nonumber \\
\mathbf{e^1} = {1 \ov k_\perp} \left(-k_y,k_x,0\right) , \quad  \mathbf{e^2} ={1 \ov k k_\perp}  \left(-k_x k_z, - k_y k_z, k_\perp^2\right) ,
\quad k = \sqrt{k_\perp^2 + k_z^2}   . \label{modes}
\end{gather}
The evolution of $\hat J^{1,2}$ is respectively governed by~\eqref{d2} and~\eqref{d0}. 

While it is not surprising that in the presence of the vector chemical potential, the diffusion processes are no longer isotropic, the equations~\eqref{d0},~\eqref{d1} and~\eqref{d2} provide precise predictions for the diffusion patterns and 
the corresponding diffusion constants. 


\subsubsection{Real time evolution in the D1 case}

It is instructive to see how an initial charge density actually diffuses when it is governed by equations such as~\eqref{d0},~\eqref{d1}  and~\eqref{d2}. Below, we consider some initial conditions to illustrate the diffusion behaviors that appear, although these may not correspond to realistic initial conditions for physical applications.  

Let us first consider the simplest case, where we set the chemical potential $\mu_i=0$. Now the time-evolution of the charge density is described by \eqref{EqDiff0}, and all spatial directions are equivalent. To analyze the temporal evolution of the initial charge density $J^{0i}(0 ,\vb{x})$, we first Fourier transform $J^{0i}$ to spatial $3$-momentum space via
\begin{align}\label{Fourier3}
J^{0i}(0,\vb{x}) = \int \frac{d^3k}{(2\pi)^3} e^{i\vb{k}\cdot\vb{x}} J^{0i}(0,\vb{k})  . 
\end{align}
Then Eq.~\eqref{EqDiff0} implies,  
\begin{align}\label{DiffusionEq0Sol}
J^{0i}(t,\vb{x}) = \int \frac{d^3k}{(2\pi)^3} e^{i\vb{k}\cdot\vb{x}} e^{-D |\vb{k}|^2 (t-t_0)} J^{0i}(0 ,\vb{k}) \equiv \int d^3 x' \, G(t, \vb{x} - \vb{x}')J^{0i}(0 ,\vb{x'})  . 
\end{align}
where the Green's function of the diffusion equation is given by the standard expression 
\begin{align}\label{Green0}
G(t, \vb{x} - \vb{x}') = \int  \frac{d^3k}{(2\pi)^3} e^{i\vb{k}\cdot(\vb{x}-\vb{x}')} e^{-D |\vb{k}|^2 t}  = \frac{1}{[4\pi D t]^{3/2} } e^{- \frac{(\vb{x}-\vb{x}')^2 }{4 D t}   } .
\end{align}
Consider for example an initial configuration of a constant charge density excited along an infinite string pointing in the $z$ direction, i.e.
\begin{align}
J^{0i} (0 , \vb{x} ) = \CJ_0 \, \delta(x) \delta(y) \delta^{iz} .
\end{align}
From \eqref{DiffusionEq0Sol}, we find that the charge density $J^{0z}$ diffuses cylindrically around the string and that the $J^{0a}$ components remain zero:
\begin{align}\label{CylDiffEq}
J^{0i} (t, \vb{x}) = \frac{\CJ_0}{4\pi D t} e^{- \frac{x^2 + y^2 }{4 Dt}  } \delta^{iz} .
\end{align}

As another example, consider an initial charge density which is composed of two semi-infinite lines meeting at an angle of $\pi/2$. We set the coordinate system so that the string of charge $J^{0x}$ runs along the positive $x$-axis and the string of $J^{0y}$ charge along the positive $y$-axis:
\begin{equation}
J^{0i} (0, \vx) = \CJ_0 \left[  \theta(x) \delta(y) \delta(z) \delta^{ix} - \delta(x) \theta(y) \delta(z) \delta^{iy} \right].
\end{equation} 
The time evolution of this $J^{0i}$ then becomes
\begin{equation}
J^{0i} (t, \vx) = \frac{ \CJ_0}{4\pi D t } \left[  F(t,x) \, e^{-\frac{y^2+z^2}{4D t}}  \delta^{ix} - F(t,y) \,e^{-\frac{x^2+z^2}{4D t} }  \delta^{iy}  \right],
\label{turning_line_later}
\end{equation} 
where $F(x,t)$ is expressed in terms of the error function:
\begin{equation}\label{Ffn}
F(t,x) = \frac{1}{2} \left[ 1+ \erf\left(\frac{x}{\sqrt{4 D t}}\right) \right] .
\end{equation} 
Note that in \eqref{turning_line_later}, the direction of the charge density at $t>0$ becomes dependent on the location in the $x$-$y$ plane, but it does not develop a non-zero $J^{0z}$ component anywhere.

Now consider the case where $\mu^i = \mu \, \delta^{iz}$, with non-zero $\mu$. To study the time evolution, we decompose the initial charge density $J^{0i} (0, \vx)$  at $t=0$ in terms of the two modes $\bfe^{1,2}$ in \eqref{modes}, i.e. 
$J^{0i} (0, \vx) = J_1^{0i} (0,\vx) +  J_2^{0i} (0, \vx)$, where
\begin{align}
J_1^{0i} (0,\vx) &= \int \frac{d^3k}{(2\pi)^3} \, e^{i \vk\cdot \vx} \lambda_{1}(\vk) \bfe_i^1(\vk)  , \label{J1}\\
J_2^{0i} (0,\vx) &= \int \frac{d^3k}{(2\pi)^3} \, e^{i \vk\cdot \vx} \lambda_{2}(\vk) \bfe_i^2(\vk ) . \label{J2}
\end{align}
The time-evolution then takes the form 
\begin{equation} 
J^{0i}(t,\vx) = \int \frac{d^3k}{(2\pi)^3} \, e^{i \vk\cdot \vx}  \le( \lambda_{1}(\vk) \bfe_i^1(\vk)  e^{- \tilde D_\perp k_\perp^2 t} + 
\lambda_{2}(\vk) \bfe_i^2(\vk )  e^{- D_\perp k_\perp^2 t } \ri)  e^{-  D_\para k_z^2 t}   \ .
\label{eqn:solution}
\end{equation} 

Let us now look at some initial configurations. 
First consider the case where the initial charge density is a Gaussian of some finite width centered around the $z$-axis and pointing in the $z$-direction, 
\begin{align}
J^{0i} ( 0 , \vb{x} ) = \CJ_0 \frac{1}{\pi \sigma^2}\, e^{- \frac{x^2+y^2}{\sigma^2}} \, \delta^{iz} ,
\label{initial_simplez}
\end{align}
For this case, $\lam_1 (\vk) =0$ and $\lam_2 (\vk) = 2 \pi \CJ_0 \de (k_z) e^{-\frac{k_x^2\sigma^2}{4}- \frac{k_y^2\sigma^2}{4}}$. 
The charge density spreads out cylindrically, 
and its direction remains constant at all points,  
\begin{align}\label{D1-String1}
J^{0i} (t, \vb{x}) = \frac{\CJ_0}{\pi(4 D_{\perp} t+ \sigma^2)} e^{- \frac{x^2 + y^2 }{4 D_\perp t + \sigma^2}  } \delta^{iz} .
\end{align}
In this case the pattern of the evolution is the same as that for $\mu =0$, although the diffusion constant is different. 

\begin{figure}[!h]
\includegraphics[width=0.95\textwidth]{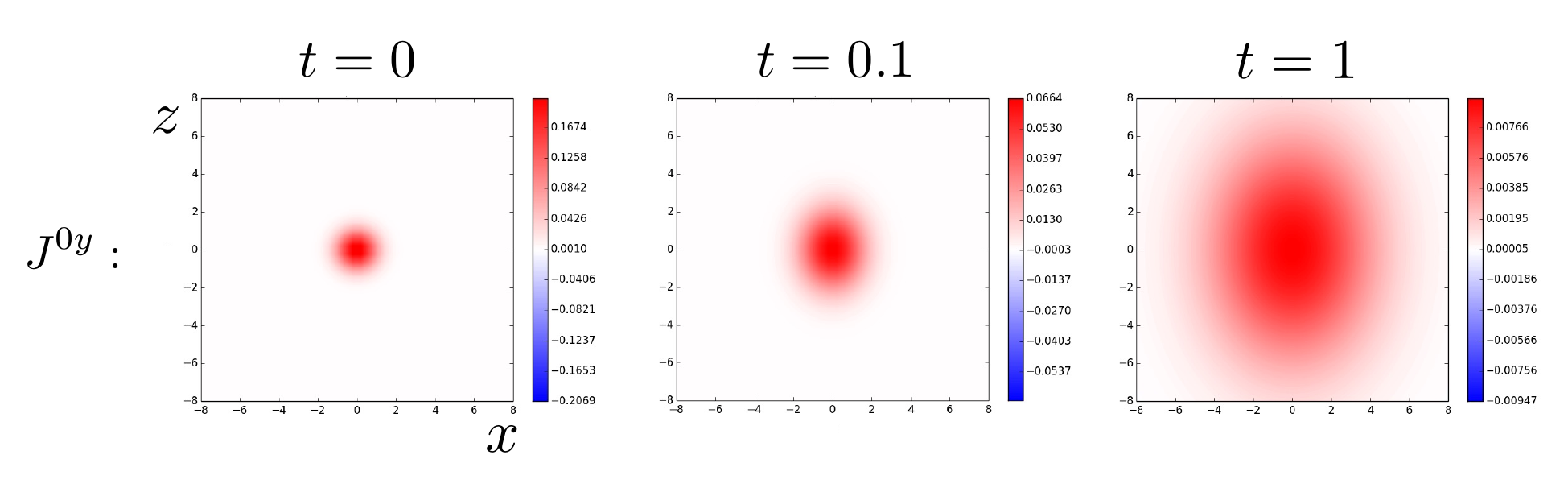}
\caption{Time-evolution of $J^{0y}$ for the initial charge density \eqref{D1Init1} in any plane orthogonal to the $y$ direction in the D1 case, with $\sigma=1$, $D_{\para}= 8$, $D_{\perp}=1$, and $\tilde D_{\perp} = 4$. $J^{0x}$ and $J^{0z}$ are zero for all times at all points.} 
\label{fig:simp_line_b}
\end{figure}

Next, let us take the initial charge density to be a Gaussian cylinder orthogonal to $\mu_i$,  
\begin{equation}\label{D1Init1}
J^{0i}(0, \vx) = \CJ_0 \frac{1}{\pi \sigma^2}\, e^{-\frac{x^2 + z^2}{\sigma^2}} \delta^{iy} ,
\end{equation}
for which $\lam_1 (\vk) = 2 \pi \CJ_0 \de (k_y)e^{-\frac{k_x^2\sigma^2}{4}-\frac{k_z^2\sigma^2}{4}}$ and $\lam_2({\bf k}) = 0$. 
The time  evolution of the charge density is 
\begin{equation}
J^{0i}(t,\vx) = \frac{\CJ_0}{ \pi \sqrt{(4 t \tilde D_\perp + \sigma^2) (4 t D_\para + \sigma^2)}} e^{-\frac{x^2}{4\tilde D_\perp t + \sigma^2}-\frac{z^2}{4D_\para t + \sigma^2}} \, \delta^{iy} .
\end{equation} 
Under time-evolution, the charge density continues to point in the $y$ direction at all times.  It spreads in an anisotropic manner, so that the surfaces of equal charge density in the $x$-$z$ plane are ellipses. See Fig.~\ref{fig:simp_line_b}. 

\begin{figure}[!h]
\includegraphics[width=0.95\textwidth]{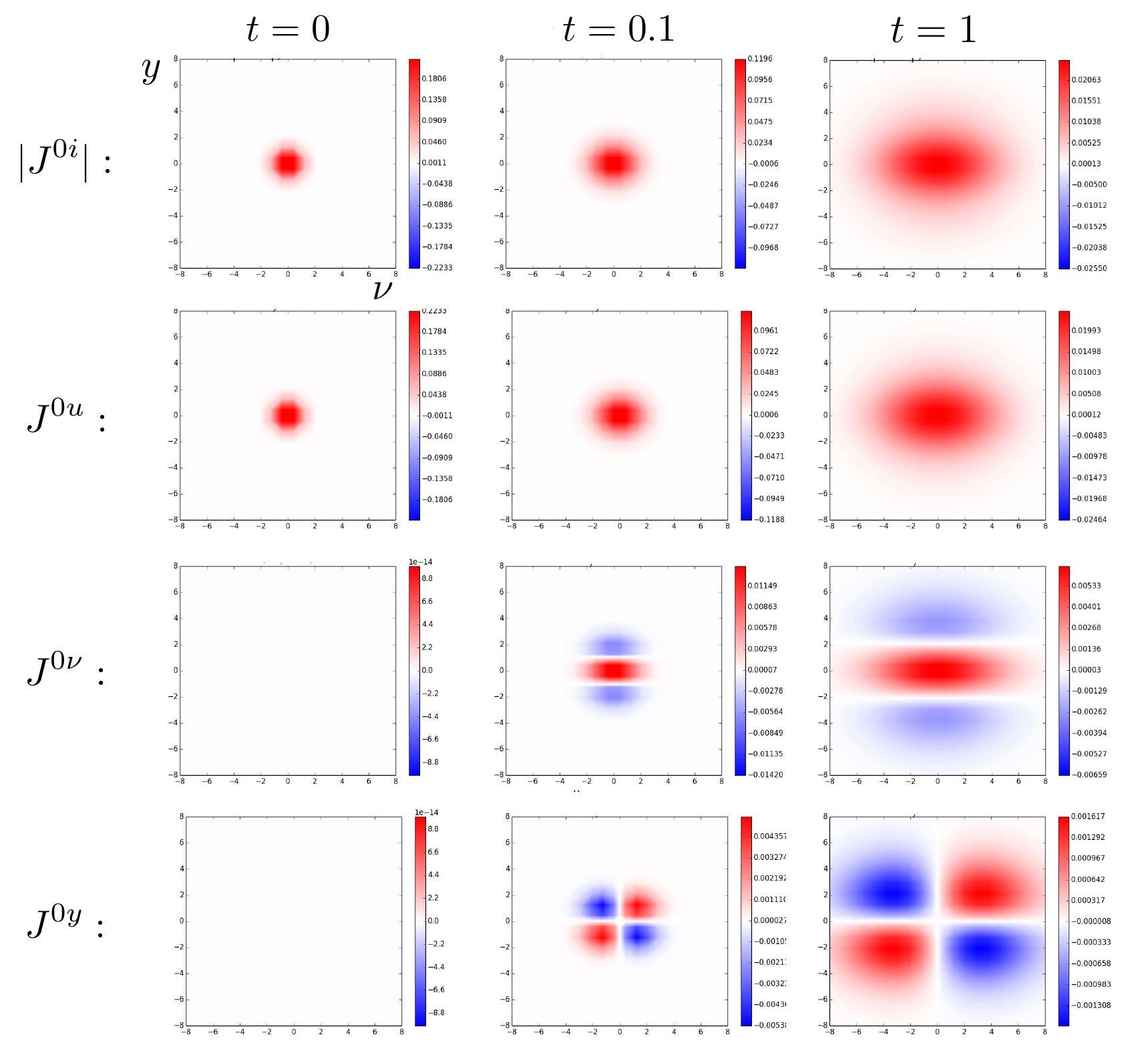}
\caption{Evolution of the magnitude and different components of $J^{0i}$ with the initial condition in \eqref{initial_tilted} in the D1 case, for any plane orthogonal to the $u$-direction in which the initial charge density points. We set the parameter values $\varphi= \pi/4$, $\sigma=1$, $D_{\para}= 8$, $D_{\perp}=1$, and $\tilde D_{\perp} = 4$.} 
\label{fig:tilted_timeev}
\end{figure}

If the charge density is neither parallel nor perpendicular to $\mu_i$,  the time evolution of $J^{0i}$ becomes significantly more complicated, exhibiting both an intricate pattern of spread and a time-dependent motion of its direction. Let us take the initial Gaussian cylinder to be centered on a line in the $x-z$ plane that passes through the origin and makes an angle $\varphi$ with the $x$-axis, and take the direction of the charge density to be along this same line, 
\begin{equation} 
J^{0i}(0, \vx) = 
\CJ_0 \, \frac{1}{\pi \sigma^2} \, e^{- \frac{v^2 + y^2}{\sigma^2}} ,  \delta^{iu}
\label{initial_tilted}
\end{equation} 
where we have changed coordinates from $x, \,  z,$ and  $y$ to $u, \,  v,$ and  $y$, with
\be 
u = x \cos \vp + z \sin \vp, \quad v = -x \sin \vp + z \cos \vp , \quad \delta^{iu} = \cos \vp \,  \delta^{ix} + \sin \vp \, \delta^{iz}  . 
\ee
In momentum space, \eqref{initial_tilted} has the form 
\be
J^{0i}(0, \vk) = 
2 \pi  \CJ_0 \, e^{ - \frac{\sigma^2}{4} (k_v^2 + k_y^2)} \, \delta(k_u) \, \delta^{iu},
\ee
which can be decomposed along the two modes ${\bf e}^{1,2}$  and evolved in time to find 
$J^{0i} (t, \vx) = J_1^{0i} (t,\vx) +  J_2^{0i} (t, \vx)$, where 
\begin{align}
J^{0i}_1 (t, \vk) &= - 2 \pi \CJ_0 \, \de (k_u) {k_y  \cos \varphi  \ov k_\perp^2} \, e^{ - \frac{\sigma^2}{4} (k_v^2 + k_y^2)} e^{- (\tilde D_\perp k_\perp^2 + D_\para k_z^2) t } \, (-k_y \delta^{ix} + k_x \delta^{iy}) ,  \label{k1}\\
 J^{0i}_2 (t, \vk) &= 2 \pi \CJ_0\,  \de (k_u)  {\sin \vp  \ov k_\perp^2} \, e^{ - \frac{\sigma^2}{4} (k_v^2 + k_y^2)} e^{- (D_\perp k_\perp^2 + D_\para k_z^2) t} \, (- k_x k_z \delta^{ix} -k_y k_z \delta^{iy} + k_\perp^2 \delta^{iz}) . \label{k2}
\end{align} 
Note that \eqref{k1} and \eqref{k2} are written in the $x$, $y$ and $z$ coordinates. On numerically Fourier-transforming back to position space, we get the time-evolution shown in Fig. \ref{fig:tilted_timeev}. 
We observe the following features: 
\begin{enumerate} 
\item The magnitude $|J^{0i}|$ of the charge density vector and its $J^{0u}$ component diffuse anisotropically in the $v-y$ plane.  
\item Although the initial charge density is entirely in the $u$-direction, $J^{0v}$ and $J^{0y}$ components are generated by the time-evolution and subsequently diffuse. These components  acquire both positive and negative values depending on the location in the $v-y$ plane. 
\end{enumerate}

\begin{figure}[!h]
\centering
\resizebox{0.52 \textwidth}{!}{\includegraphics{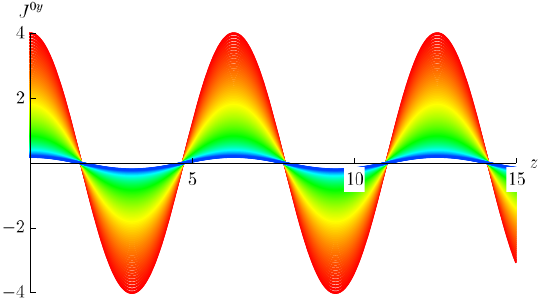}}
\caption{Evolution of the initial condition \eqref{D2ExTime0} in the D1 case. We show the dependence of the charge density component $J^{0y}$ as a function of time along the $z$-axis in units of $D = 1/4$, $\CJ_0 = 4$ and $\CK = 1$. Different instances of time (normalized by $4 D_{\parallel} \CK^2 = 1$) are plotted as different colors from red at $t=0$ to blue at $t = 12$. $J^{0x}$ and $J^{0z}$ remain zero at all points and all times.}
\label{fig:D1Diffusion}
\end{figure}

As another example of the initial charge density, which will be useful for comparison with diffusion phenomena coming from other effective actions with various discrete symmetry violations, we can consider a configuration which is extended over all space, points in the $y$ direction everywhere, and has an oscillatory dependence on the $z$ coordinate:    
\begin{align}\label{D2ExTime0}
J^{0i}(0,\vx)=  \CJ_0 \, \cos (\CK z) \,  \delta^{iy} .
\end{align}
Such an initial charge density simply decays in magnitude exponentially at all points, with no change in its direction or the shape of its spatial distribution: 
\begin{align}\label{D1Ex}
J^{0i}(t,\vx) & = \CJ_0\, e^{- D_{\para} \CK^2 t} \cos (\CK z) \delta^{iy} . 
\end{align}
This time-evolution is shown in Fig. \ref{fig:D1Diffusion}. 

\subsection{Equations of motion in D2 case}

Let us now understand the diffusion behaviour in the effective action $\sL^{(2)}_{\rm D2}$. The results in this subsection were previously summarized in \cite{Vardhan:2022wxz}, where it was noted that $\sL^{(2)}_{\rm D2}$ describes magnetohydrodynamics in neutron stars. In this case, after expanding the constitutive relations to linear order in $f_{r0i}$, we find new terms in the current and charge densities in addition to those in the $J_0^{ij}$ from \eqref{j1}--\eqref{JC-4-2}:
\begin{align} \label{hsb2}
J_{\rm D1}^{0i} &= J_{\rm D1}^{0i}  , \quad  \hat J_{\rm D2}^z = {\hat{J}_{\rm D2}}^{z} ,\\
J_{\rm D2}^{za} &=J_{\rm D1}^{za}  + p \mu \ep_{ab} \p_0 G_{rzb}  . 
\label{hsb3}
\end{align}
In the rest of this subsection we drop the explicit D2 subscript. 
The susceptibilities $\chi_{\parallel}$ and $\chi_{\perp}$ are the same as in the D1 case. However, for the conductivities we now find 
\be 
\sigma^{\perp}_{ab} \equiv \sigma^\perp_1 \delta_{ab} - \sigma_2^\perp \epsilon_{ab}=  2 \beta_0 d \de_{ab} - p \mu \ep_{ab} , \quad \sigma^{\para} = 2\beta_0(d+2 \tilde d \mu^2) ,
\label{conductivities3}
\ee
with $\sig^{\perp}_{ab}$ having a new Hall-like term, which leads to a current density $J^{zx} \, (J^{zy})$ in response to a higher-form electric field $E^{zy}\, (E^{zx})$. This term involving $\sigma_2^{\perp}$ will lead to differences in the dispersion relations and the equations of motion at quadratic order in momentum relative to the D1 case. 

\subsubsection{Diffusion equations}

Let us define $\chi_{\perp}$, $\chi_{\para}$, $\sigma_1^{\perp}$ and $\sigma^{\para}$ as in the D1 case, and 
\be 
\sigma_2^{\perp} \equiv - p \mu . 
\ee
Then the relevant parts of $J^{\mu\nu}$ for understanding the equations of motion to order $k^3$ are
\begin{align}
J^{0z} & =a \mu +  \chi_\parallel f_{r0z} , \quad J^{0a} = \chi_\perp f_{r0a} \label{450}, \\
J^{za} &= - \sig^\perp_1 (\p_z f_{r0a} - \p_a f_{r0z}) - \sig^\perp_2 \ep_{ab}  (\p_z f_{r0b} - \p_b f_{r0z}) , \\
\hat J^z &= - \sig^\para \ep_{ab} \p_a f_{r0b}   .
\end{align} 
where we have used~\eqref{FieldRel1} in $J^{za}$.   
The equations of motion then become 
\begin{gather}\label{e1}
 \p_0 J^{0z} -\sig^{\perp}_1 \le[{1 \ov \chi_\perp} \p_z^2 + {1 \ov \chi_\para} \p_a^2 \ri]J^{0z}  + {\sig_2^\perp \ov \chi_\perp} 
 \ep_{ab} \p_a \p_z J^{0b} = 0,    \\
\!\! \p_0 J^{0a} -  \le[ \frac{\sig^\perp_1}{\chi_\perp}  \p_z^2  + \frac{\sig^\para}{\chi_\perp} \p_b^2 \ri] J^{0a} + 
 \le({\sig^\para \ov \chi_\perp} - {\sig^\perp_1 \ov \chi_\para} \ri) \p_a \p_b J^{0b} - \sig_2^\perp \ep_{ab} 
\le[{\p_z^2 \ov \chi_\perp} J^{0b} + {\p_b \p_c \ov \chi_\para} J^{0c} \ri] \!  = 0 . 
\label{e2}
 \end{gather}  All three equations are now coupled. However, there should again be only two independent modes due to the requirement that the charge density is divergenceless. Indeed, acting with $\p_z$ on~\eqref{e1} and acting with $\ep_{ca} \p_c $ on~\eqref{e2}, we find that 
\begin{align} \label{DD1}
(\p_0 + \CD ) \bma Q_1 \cr Q_2 \ema = 0 ,  
\quad \CD =  \bma D_{11}  & D_{12} \cr D_{21} & D_{22} \ema ,
\end{align}
where 
\begin{gather}
Q_2 \equiv \ep_{ca} \p_c  J^{0a}, \quad
Q_1 \equiv \p_a J^{0a} = - \p_z J^{0z} , \\
D_{11} =  - \sig^{\perp}_1 \le[ {1 \ov \chi_\perp} \p_z^2 + {1 \ov \chi_\para} \p_a^2 \ri]
=   \sig^{\perp}_1 \le[ {1 \ov \chi_\perp} k_z^2 + {1 \ov \chi_\para} k_a^2 \ri] , \quad
D_{12} =  -{\sig_2^\perp \ov \chi_\perp}
 \p_z^2  =  {\sig_2^\perp \ov \chi_\perp} k_z^2 \\
D_{21} =   \sig_2^\perp \le({\p_z^2 \ov \chi_\perp} + {\p_a^2 \ov \chi_\para} \ri)   = -\sig_2^\perp \le({k_z^2 \ov \chi_\perp} + {k_a^2 \ov \chi_\para} \ri) , \\
D_{22} = - {1 \ov \chi_\perp} \le[ \sig^\perp_1  \p_z^2  + \sig^\para \p_b^2 \ri] = 
{1 \ov \chi_\perp} \le[ \sig^\perp_1  k_z^2  + \sig^\para k_b^2 \ri]   .
\end{gather}
Requiring that the determinant of the coefficient matrix of the system of equations \eqref{DD1} vanishes, we obtain the dispersion relation 
\bea 
\om &= &  - i {D_{11} + D_{22} \pm \sqrt{(D_{11} - D_{22})^2 + 4 D_{12} D_{21}} \ov 2}  \label{fin}\\  
& = &  - i  {\sig^{\perp}_1 \ov \chi_\perp} k_z^2 -  \frac{i}{2} \le({\sig^\para \ov \chi_\perp} + {\sig^{\perp}_1 \ov \chi_\para}   \ri) k_a^2  
\mp \frac{i}{2} \sqrt{    \le({\sig^\para \ov \chi_\perp} - {\sig^{\perp}_1 \ov \chi_\para}   \ri)^2 (k_a^2)^2   - 4  \left(   \frac{ \sig^\perp_2}{  \chi_\perp} \right)^2 \le(k_z^2 + \frac{\chi_\perp}{\chi_\para} k_a^2  \ri)  
 k_z^2  }   .  \nonumber
\eea
The quantity inside the square root of~\eqref{fin} can in principle be negative, and thus the right hand side of~\eqref{fin} can contain a real part.

\subsubsection{Real time evolution in the D2 case}

Similar to the discussion around \eqref{J1} and \eqref{J2}, we can decompose a general initial charge density in momentum space along two modes ${\bf \tilde{e}}^1$ and ${\bf \tilde{e}}^2$, which can be deduced from the eigenvectors of $\CD$. The dispersion relations $\omega_{1,2}({\bf k})$ for these modes are given by the two values in \eqref{fin}. 

For an initial condition pointing entirely in the $\mu_i$ direction, the new contribution to the equations of motion from $\sigma_2^{\perp}$ does not lead to any new effects, and we still see isotropic time-evolution without any change in the direction of the charge density.

\begin{figure}[!h]
\includegraphics[width=0.95\textwidth]{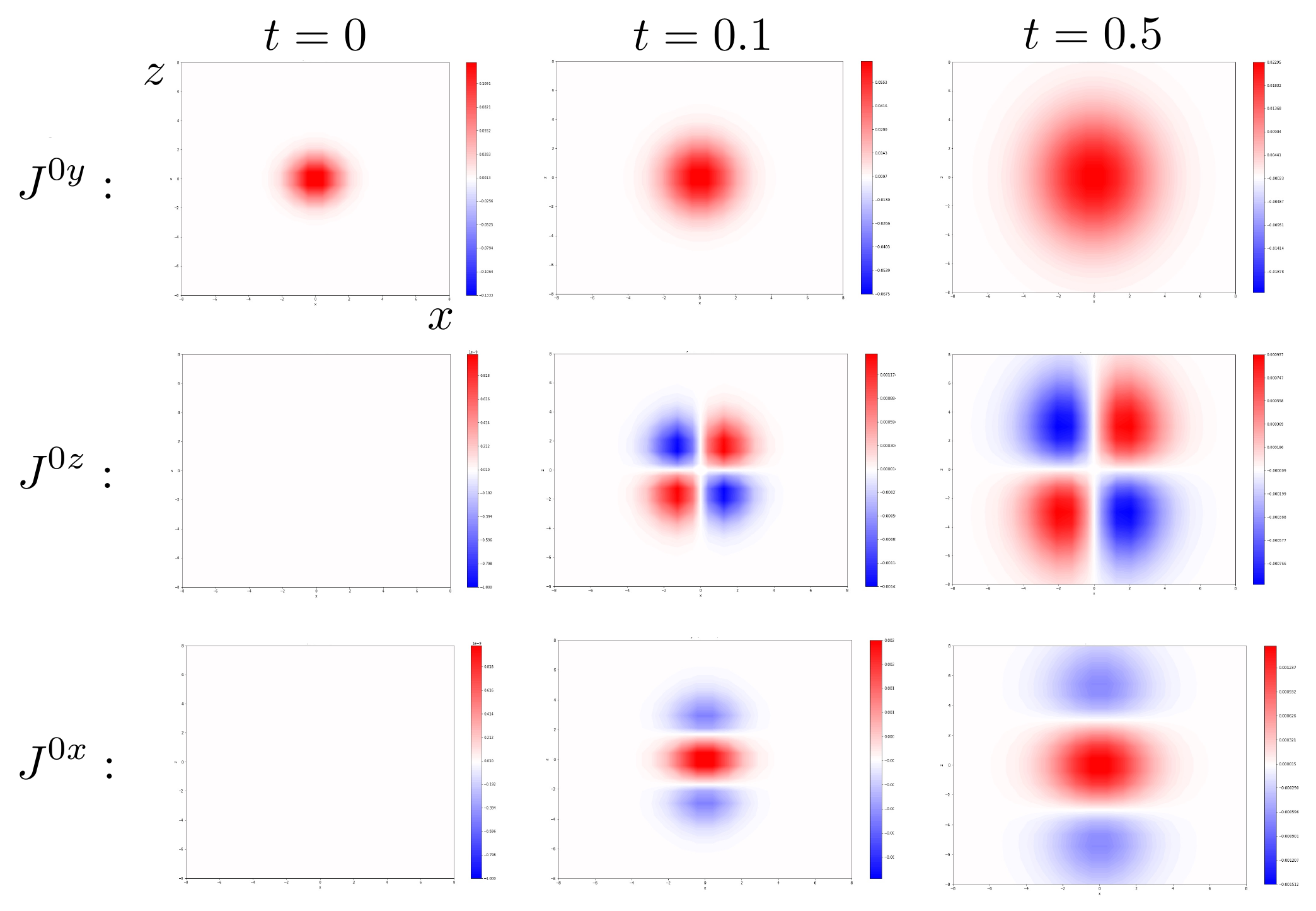}
\caption{Evolution of the different components of $J^{0i}$ with the initial condition in \eqref{D1Init1} in the D2 case, with a Gaussian density centered on the $y$ axis and pointing in the $y$ direction. This should be compared to the simple evolution for the same initial condition in Fig. \ref{fig:simp_line_b} in the D1 case.}
\label{fig:y_with_t5}
\end{figure}

If we consider an initial condition with a Gaussian cylinder along the $y$-axis and pointing in the $y$-direction as in \eqref{D1Init1}, we now see qualitatively new effects in the time-evolution compared to the D1 case. Unlike in the evolution for that case, shown in Fig. \ref{fig:simp_line_b}, we now find in the presence of $\sigma_2^{\para}$ that $J^{0x}$ and $J^{0z}$ both develop non-zero values under time-evolution and subsequently diffuse. See Fig. \ref{fig:y_with_t5}.  

\begin{figure}[!h]
\centering
\resizebox{0.55 \textwidth}{!}{\includegraphics{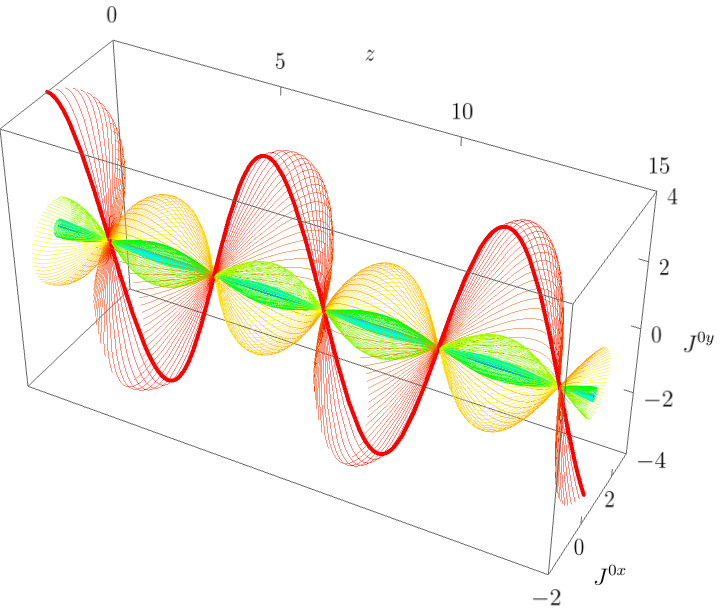}}
\hspace{0.05 \textwidth}
\resizebox{0.35 \textwidth}{!}{\includegraphics{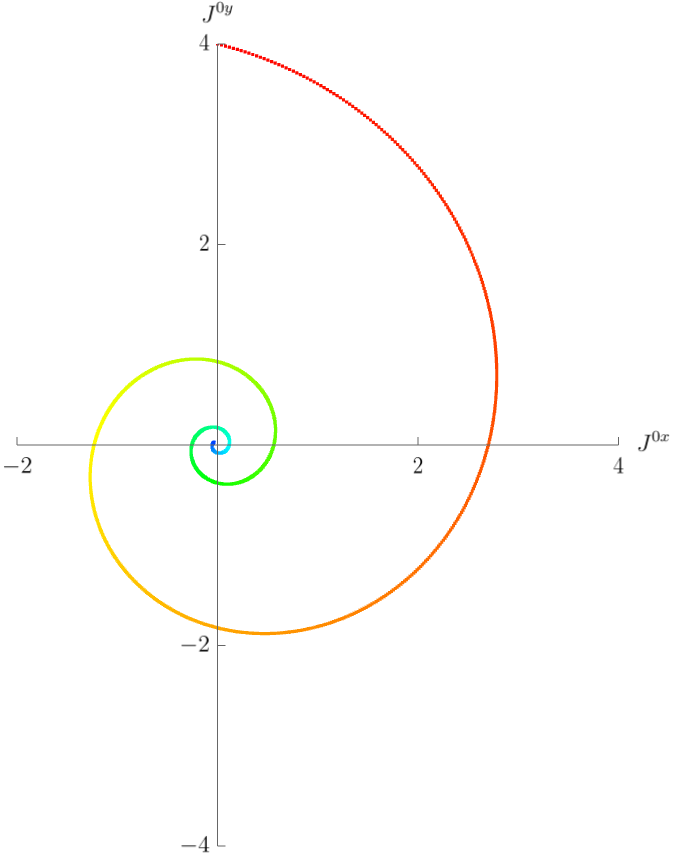}}
\caption{(Left) Evolution of the initial condition in \eqref{D2ExTime0} in the D2 case, as dictated by Eq.~\eqref{D2_DiffPlots}, with the choice of parameters $\sigma_2^{\perp}/\chi_{\perp}= -4 D_{\parallel}$ and in units of $D_\para = 1/4$, $\CJ_0 = 4$ and $\CK = 1$. We plot the dependence of the charge density components $J^{0x}$ and $J^{0y}$  along the $z$-axis. Different instances of time are plotted as different colors, organized according to hue --- from red at $t=0$ through yellow, green and towards blue at $t = 12$. (Right) The components $J^{0x}$ and $J^{0y}$ plotted at $z= 0$ for different times, which are represented by the same color coding as in the left panel. At $t=0$, the charge density is aligned with the $y$-axis (cf. Eq.~\eqref{D2ExTime0}).}
\label{fig:D2Diffusion}
\end{figure}

\begin{figure}[!h]
\centering
\resizebox{0.5\textwidth}{!}{\includegraphics{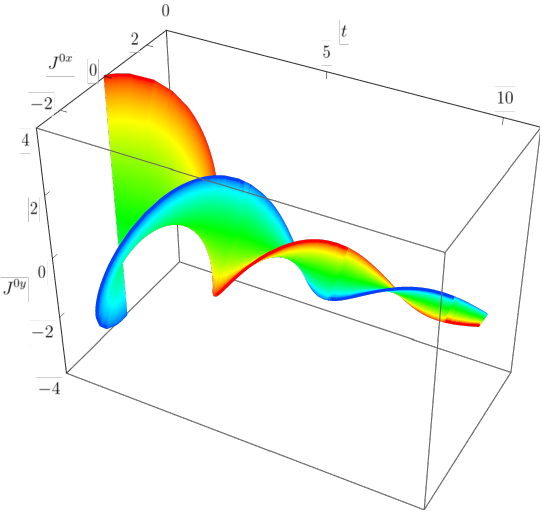}}
\caption{Decaying circularly polarized $J^{0i}$ in the D2 case is plotted for the same parameter choice and units as in Figure \ref{fig:D2Diffusion}. Different values of $z$ are plotted as different colors for half a period of $z \in [0,\pi]$ with hue running from red at $z = 0$ to blue at $z = \pi$.}
\label{fig:D2Diffusion-time}
\end{figure}

As another example with a striking difference from the case with all discrete symmetries preserved, consider the initial charge density pointing in the $y$-direction and with a cosine dependence on the $z$ coordinate, as in \eqref{D1Ex}. Now since $J^{0a}$ is dependent only on $z$ and $J^{0z} =0$, the dispersion relation from \eqref{DD1} simplifies, and we have
\be\label{C3kz01}
\om = -i {\sig^{\perp}_1 \pm i \sig^\perp_2 \ov \chi_\perp} k_z^2 ,
\ee
which can also be obtained by setting $k_a \to 0$ in~\eqref{fin}. The corresponding modes in this case also become simple, and are given respectively by 
\be 
{\bf \tilde{e}^{1}}_i =(1,  i , 0)  , \quad {\bf \tilde{e}^{2}}_i =  (1, - i , 0) . \label{simple_modes}
\ee
Decomposing the initial condition \eqref{D1Ex} along these modes and evolving in time, we see 
a ``circular diffusion'' pattern with a complex diffusion constant, 
\be\label{D2_DiffPlots}
J^{0i}(t , \vx) = \CJ_0 \cos(\CK z) \,  e^{-D_{\para} \CK^2 t} \, \left[-\sin\left(\frac{\sigma_2^{\perp}}{\chi_{\perp}} \CK^2 t \right)\, \delta^{ix} + \cos \left(\frac{\sigma_2^{\perp}}{\chi_{\perp}} \CK^2 t \right)\, \delta^{iy} \right]  .  
\ee
This time evolution is shown in Fig.~\ref{fig:D2Diffusion} and~\ref{fig:D2Diffusion-time}, and should be contrasted with the simple exponentially decaying behaviour in Fig.~\ref{fig:D1Diffusion} for the theory with all discrete symmetries preserved,  where we see no change in the direction of the charge density. 

\section{Corrections to diffusion at cubic order in momenta} 
\label{sec:weight_three}

So far, we have analysed the two possible types of diffusion behaviours resulting from the effective actions up to weight 2. In this section, we will discuss the weight 3 terms that are relevant for the linear response theory. Like in the discussion at the beginning of Section \ref{sec:EFT-Sym}, we will consider the equations of motion up to linear order in $f_{r0i}$ and set $a$-fields to zero in the constitutive relations. This means that in the most general effective action \eqref{l3gen} of Appendix \ref{App:KMS} (before assuming any particular choice of discrete symmetries), we can ignore the $C, Q, R, S, T$ terms, and consider
\begin{align}~\label{51} 
\sL^{(\text{linear})} =  
\sL^{(2)} &+ F_{ijk} G_{a0i} \partial_j G_{r0k}  \\ 
& + M_{ijk} G_{a0i} G_{ajk} +   N_{ijk} G_{ajk} \partial_0 G_{r0i}  +  O_{ijk} G_{a0i} \partial_0 G_{rjk}    \label{52}
\\
&+ U_{ijklm}  G_{aij} \partial_k G_{alm} + V_{ijklm} G_{aij} \partial_k \partial_0 G_{rlm} + W_{ijklm} \partial_0 G_{rij} \partial_k G_{alm} . \label{53} 
\end{align} 
In all cases, the KMS conditions will determine the tensors $U$ and $W$ in terms of $V$, and $M$ and $N$ in terms of $O$. The precise relations, and whether some of the tensors (or their even and odd parts under $G_{r0i} \to - G_{r0i}$) get set to zero, will depend on the choice of discrete symmetries. 

The general correction to the constitutive relations from \eqref{51}--\eqref{53} takes the form 
\begin{align} 
&J^{0i} =J^{0i}_{(2)}  + (F_{ijk} + 2 O_{i[jk]})\partial_j G_{r0k} ,\label{54} \\
& J^{ij}  = J^{ij}_{(2)} + 2 N_{k[ij]} \partial_0 G_{r0k} + 4 (V_{[ij]k[lm]}-W_{[lm]k[ij]})\partial_k \partial_l G_{r0m} ,\label{55}
\end{align} 
where $J^{\mu \nu}_{(2)}$ refers to the constitutive relation from the weight 2 terms in the action, and we have used \eqref{FieldRel1} and kept only linear in $f_{r0i}$ terms.

Let us now write down a general expansion for each of the independent tensors $F, O, V$ appearing above, separating the terms which are even and odd under $G_{r0i} \to - G_{r0i}$. (Note that for instance $F=F^{(e)}+ F^{(o)}$.) Below all small alphabets other than the $k_i$ will refer to arbitrary real functions of $G_{r0i}^2$, and the $k_i$ are arbitrary real constants. We use the notation $\tilde \epsilon_{ij} = \epsilon_{ijk}G_{r0k}$.

Due to the KMS conditions in all cases, $F_{ijk}$ is antisymmetric under exchange of $i$ and $k$. The even and odd parts of the most general tensor of this kind are  
\begin{align} 
& F_{ijk}^{(e)} = k_1(- \tilde \epsilon_{ij} G_{r0k} + \tilde \epsilon_{kj} G_{r0i}) + k_2 G_{r0j} \tilde \epsilon_{ik} \label{f_odd}, \\
& F_{ijk}^{(o)} = k_3 (G_{r0i} \delta_{jk} - G_{r0k} \delta_{ji}).  \label{f_even} 
\end{align} 
Next, consider the terms in \eqref{52}. We can expand 
\begin{align} 
& O_{ijk}^{(o)} = l_1 G_{r0j} \delta_{ik} \label{o_odd}, \\
& O_{ijk}^{(e)} = l_2 G_{r0i} \tilde \epsilon_{jk} + l_3 \epsilon_{ijk} . \label{o_even}
\end{align} 
The same set of tensor structures appear in $N_{ijk}^{(o), (e)}$ and $M_{ijk}^{(o), (e)}$, with coefficients that are related to those in $O$ in a way that depends on the KMS transformations. Finally, consider the last line of \eqref{53}. We have 
\begin{align}\label{v_odd}
V_{ijklm}^{(o)}= &\,\, v_1 \tilde\epsilon_{ik} \epsilon_{jlm} 
+   \left(   v_2 \delta_{jk}\delta_{lm} + v_3 \delta_{jm}\delta_{kl}  \right) G_{r0i} + \left (  v_4\delta_{il}\delta_{jm} +v_{5}\tilde\epsilon_{ij}\tilde\epsilon_{lm} \right)  G_{r0k}   \nn
&+  v_{6}G_{r0i} G_{r0k} G_{r0l} \delta_{jm} + v_{7}G_{r0i} G_{r0l} G_{r0m} \delta_{jk} ,
\end{align}
and 
\begin{align}  \label{v_even}
V_{ijklm}^{(e)} =&\,\,  v_8 \delta_{kl} \epsilon_{ijm}+ v_9 \delta_{km} \epsilon_{ijl} + (v_{10}  \delta_{jk}\tilde\epsilon_{lm} + v_{11} \tilde\epsilon_{jk} \delta_{lm} + v_{12}\tilde\epsilon_{jm} \delta_{kl}) G_{r0i} +    v_{13} \tilde\epsilon_{ij} \delta_{kl}G_{r0m} \nn
&+   v_{14}G_{r0i} G_{r0k} \epsilon_{jlm} + v_{15} G_{r0k} G_{r0l} \epsilon_{ijm} + v_{16} G_{r0k} G_{r0m} \epsilon_{ijl} +  v_{17}G_{r0i} G_{r0k} G_{r0l} \tilde\epsilon_{jm}  .
\end{align} 
$U$ and $W$ have a similar structure, and the coefficients are related to those appearing in $V$.  

By relating various terms above or setting them to zero according to the KMS conditions given in Appendix \ref{App:KMS}, \eqref{f_odd}--\eqref{v_even} together with the constitutive relations \eqref{54}--\eqref{55} can be used to obtain dispersion relations $\omega(k)$ up to cubic order in $k$ for the different symmetry classes. More explicitly, we can invert the constitutive relation for  $J^{0i}$ coming from \eqref{51}--\eqref{53} to obtain an expression for $f_{r0i}$ in terms of $J^{0i}$ up to first order in spatial derivatives, generalizing \eqref{450}. By substituting this expression for $f_{r0i}$ into the constitutive relation for $J^{ij}$ and using the current conservation equation \eqref{current_cons}, we get an equation involving various derivatives of $J^{0i}$ up to weight 3, which can be used to derive $\omega(k)$. The resulting expressions are complicated in general and we do not write them down explicitly here, but they can be simply worked out using this procedure for any particular application.  Note that since there are no weight 3 terms in $\sL_0$, $\sL_{\CP_-, \CC\CT_-}$, and $\sL_{\CP_-, \CT_-}$, the dispersion relations in these cases will not be corrected at cubic order.  In the remaining cases, some of the tensors appearing in the above expressions can drop out of the dispersion relations due to cancellations between different terms in the equation of motion, even if they naively appear in the effective action and constitutive relation.

\section{The symmetry-broken phase with relativistic invariance: the Maxwell theory of electrodynamics}\label{sec:Maxwell}

Let us now turn to the setup where the one-form symmetry in the fluid is spontaneously broken. As explained in \cite{Crossley:2015evo}, the effective field theory for a fluid with a spontaneously broken symmetry can be formulated in terms of the same set of dynamical fields that appear in the symmetry preserving phase, with the difference that we no longer impose the diagonal gauge symmetry which allowed the different fluid elements to transform independently. In the one-form context, this means that we should write down an effective action in terms of the fields \eqref{BDef}, but without imposing \eqref{1formT-B1}--\eqref{1formT-B3}. Anticipating that we will set the external sources $b_{\mu \nu}$ to zero so that $G_{r\mu \nu}$ and $G_{a \mu \nu}$ are both small, let us write down the ``ar'' part of the effective action to quadratic order in $G_{\mu \nu}$ and zeroth order in derivatives: 
\begin{align}
\CL_{SB} &=  a G_{a0i} G_{r0i} + b \epsilon_{ijk} G_{aij} G_{r0k}+ c \epsilon_{ijk}  G_{a0i} G_{rjk} + d  G_{aij} G_{rij}  .
\end{align} 
If we further impose Lorentz-invariance, we must set $d = - a /2 $ and $c= - b$. Then we get 
\begin{align} 
\CL_{SB} &= - \frac{a}{2} G_{a\mu\nu} G_r^{\mu\nu} + \frac{b}{2} \epsilon_{\mu\nu\rho\sigma} G_a^{\mu\nu} G_r^{\rho\sigma} ,\label{L0-SB-Red} \\
&=   - \frac{a}{2} G_{a\mu\nu} G_r^{\mu\nu} + b~G_{a\mu\nu} \tilde G_r^{\mu\nu},
\end{align} 
where we have used the flat Minkowski metric with the mostly positive signature, and $\tilde G_{r\mu\nu}$ is the Hodge dual of $G_{r\mu\nu}$,
\begin{align}
\tilde G_{r\mu\nu} \equiv \frac{1}{2} \epsilon_{\mu\nu\rho\sigma} G_r^{\rho\sigma} . 
\end{align}
This leads to the following two-form current:
\begin{align}
J^{\mu\nu} = - a G_r^{\mu\nu}  + 2  b \tilde G_r^{\mu\nu} .
\end{align}

The relativistic effective theory with a broken one-form symmetry, expanded to leading order in derivatives, is thus precisely the Maxwell theory of free photons (see also Refs.~\cite{Lake:2018dqm,Hofman:2018lfz,iqbal_mcgreevy}). Note that, so far, we have not imposed any discrete symmetries, which is why the current has two independent pieces. In the standard language, the effective action is  
\begin{align}\label{MaxwellAction}
S = \int d^4 x \left( - \frac{1}{4e^2} F_{\mu\nu} F^{\mu\nu} + \frac{1}{2g^2} F_{\mu\nu} \tilde F^{\mu\nu}   \right) ,
\end{align}
where $F_{\mu\nu} = \partial_\mu a_\nu - \partial_\nu a_\mu$ is the electromagnetic field strength and its Hodge dual is $\tilde F^{\mu\nu} = \frac{1}{2} \epsilon^{\mu\nu\rho\sigma} F_{\rho\sigma}$. In terms of our effective action, we can either choose to identify
\begin{align}
G_{\mu\nu} = F_{\mu\nu}\quad \text{or} \quad G_{\mu\nu} = \tilde F_{\mu\nu}.
\end{align}
Using the definition \eqref{BDef} in the absence of an external two-form $b_{\mu\nu}$, we thus conclude that the Stueckelberg field $A_\mu$ is either the photon $a_\mu$, or by writing $\tilde F_{\mu\nu} = \partial_\mu \tilde a_\nu - \partial_\nu \tilde a_\mu$, the magnetic photon $\tilde a_\mu$. The hydrodynamic gapless mode has become the Goldstone mode of the spontaneously broken one-form symmetry: the photon. This is precisely analogous to the situation with a spontaneously broken zero-form $U(1)$ symmetry discussed in \cite{Crossley:2015evo}, where the symmetry-broken phase is a superfluid.

\section{Conclusions and Discussion}\label{sec:Disc}

In this paper, we provided a general classification of the effective actions for hydrodynamic modes associated with a one-form $U(1)$ symmetry, in a probe limit where the dynamics of the one-form charge density is decoupled from that of the energy-momentum tensor. We studied the two possible types of diffusion that can be exhibited by such a system depending on whether or not it involves a charge conjugation symmetry, deriving both the dispersion relations and the real-time evolution of various initial configurations. We presented the terms that can give corrections to the dispersion relations at cubic order, which take different forms depending on the precise discrete symmetries present in the system such as parity, time-reversal, and charge conjugation. We also derived the effective action in the case where the one-form symmetry is spontaneously broken, and showed that it reduces to the Maxwell action, providing an explicit realization of the idea that the photon is the Goldstone boson associated with spontaneous breaking of a one-form symmetry. 

One physical application of these effective actions, which we discussed in detail in~\cite{Vardhan:2022wxz}, is to the case of neutron stars. An interesting future direction would be to find other concrete physical realizations of the various discrete symmetry classes discussed in this paper, for example in various exotic condensed matter systems. While the weight 2 terms in the effective actions fell into just two cases for all discrete symmetry cases, the weight 3 terms can further distinguish the hydrodynamic behaviours for the different discrete symmetries at next-to-leading order. It would be interesting to better understand the physical consequences of the various weight 3 terms, both for the dispersion relations and for the real-time evolution of initial charge densities. 

\begin{acknowledgments}
The work of SG was supported by the STFC Ernest Rutherford Fellowship ST/T00388X/1. The work is also supported by the research programme P1-0402 and the project N1-0245 of Slovenian Research Agency (ARIS). SV is supported by Google. 
HL is supported by the Office of High Energy Physics of U.S. Department of Energy under grant Contract Number  DE-SC0012567 and DE-SC0020360 (MIT contract \# 578218).
\end{acknowledgments}

\appendix

\section{Deriving the effective actions}\label{App:KMS}

As discussed in the main text, when doing the derivative counting, we should assign weight $2$ to $\p_0$,  weight $1$ to $\p_i$, and weight $0$ to $G_{r0i}$.  Under this assignment, $G_{rij}, G_{a0i}, G_{aij}$ have respective weights $-1, 2, 1$. Recall that each term in the action must have at least one $a$-field, and that $G_{rij}$ can appear either as $\partial_0 G_{rij}$, which has weight 1, or $H_{rijk}$, which has weight 0. $H_{rijk}$ can therefore appear an arbitrary number of times. For the purpose of finding the dispersion relations, we can set external sources to zero, so we will not explicitly write down terms involving $H_{r \lambda \mu \nu }$. Recall that in the absence of external sources, 
\be 
\partial_0 G_{rij} = 2 \partial_{[i} G_{r0j]} , \quad H_{0ijk}=0 . 
\ee
The form on the LHS will be more useful for the purpose of imposing the dynamical KMS relations, so we will not write down explicit terms involving  $\partial_{[i} G_{r0j]}$ in the action. Note that we should explicitly include terms involving the symmetric combination $\partial_{(i} G_{r0j)}$. 

In the expressions below, tensors such as $B_{ij}$, $A_{i}$, $D_{ijkl}$, and so on are all constructed from arbitrary combinations of $G_{r0i}$, $\delta_{ij}$, and $\epsilon_{ijk}$ and have weight zero. 
We can see under the above constraints that the only term of weight 1 that can appear in the action is 
\be  \label{action_1}
\sL^{(1)} = B_{ij} G_{aij}.
\ee
The terms of weight 2 are: 
\begin{align} \label{action_2}
\sL^{(2)} =&~  A_i G_{a0i}   + D_{ijkl} G_{aij} G_{akl}  + P_{ijkl} G_{aij} \partial_0 G_{rkl}  +   E_{ijkl}G_{aij} \partial_{(k} G_{r0l)} .
\end{align} 
The terms of weight 3 are: 
\begin{align} \label{action_3}
\sL^{(3)} = &~   F_{ijk} G_{a0i} \partial_j G_{r0k}  \nn 
& +  K_{ijklm} G_{aij}   \partial_{k} \partial_{(l} G_{r0m)} \nn 
&
+ L_{ijklmn} G_{aij}   \partial_{(k}  G_{r0m)} \partial_{(l} G_{r0n)} \nn 
& + M_{ijk} G_{a0i} G_{ajk} +   N_{ijk} G_{ajk} \partial_0 G_{r0i}  +  O_{ijk} G_{a0i} \partial_0 G_{rjk}   
\nn
 & + U_{ijklm}  G_{aij} \partial_k G_{alm} + V_{ijklm} G_{aij} \partial_k \partial_0 G_{rlm} +  W_{ijklm} \partial_0 G_{rij} \partial_k G_{alm}  \nn 
 & + C_{ijklmn} G_{aij}  G_{akl} \partial_{(m} G_{r0n)}  
+Q_{ijklmn}  G_{aij} \partial_0 G_{rkl} \partial_{(m} G_{r0n)}  \nn 
 & +  R_{ijklmn}G_{aij} G_{akl} G_{amn} +  S_{ijklmn}G_{aij} G_{akl} \partial_0 G_{rmn} + T_{ijklmn}  G_{aij} \partial_0 G_{rkl} \partial_0 G_{rmn}.
\end{align} 

Let us now impose invariance under the various dynamical KMS transformations to the above actions. For this purpose, it is useful to  divide each of the above coefficient tensors into a part which is even and a part which is odd under $G_{r0i} \to - G_{r0i}$. We will label these parts with $(e)$ or $(o)$ superscripts. We will see below that  the only terms that can be written down in $A_{ij}$ and $B_i$ are odd, and $D_{ijkl}$ is purely even, but all other coefficients can in principle have both even and odd parts before imposing the dynamical KMS conditions.  In \eqref{action_1}--\eqref{action_3}, we have written all terms whose KMS transformations can cancel with each other in a single line. For the terms which appear by themselves in a single line (i.e., they cannot combine with other terms to satisfy the KMS conditions), the part which changes sign under the KMS transformation (which can be either the even or the odd part depending on the symmetry) must be set to zero. The remaining terms, which do not change sign, can be non-zero if there is a way for the change under the KMS transformation to combine into a total derivative of the form $\partial_{\mu} X^{\mu}$ for some vector $X^{\mu}$. Note that the $W$ term is related to the $Q$ and $T$ terms by integration by parts, but it is useful to write it separately for the purpose of imposing the KMS conditions.

Let us first discuss the KMS conditions which apply in all four cases~\eqref{CaseIKMS}--\eqref{CaseIIKMS}. In the points below, all coefficients labelled by small alphabets can be arbitrary functions of $G_{r0i}^2$, and we introduce the notation $\tilde \epsilon_{ij} = \epsilon_{ijk} G_{r0k}$. 
\begin{enumerate} 
\item $B_{ij}$ can be expanded in the most general case to 
\be
B_{ij} = b \tilde \epsilon_{ij}    . 
\ee
Under $\text{KMS}_I$ and $\text{KMS}_{II}$, this term changes sign, and hence must be set to zero. Under $\text{KMS}_{III}$ and $\text{KMS}_{IV}$, 
\be 
b \epsilon_{ijk} G_{r0k} G_{aij} \rightarrow b \epsilon_{ijk} G_{r0k} G_{aij} + b \epsilon_{ijk} G_{r0k} \partial_0 G_{rij} .
\ee
The second term cannot be written in the form $\partial_{\mu}X^{\mu}$, so we set $b=0$ in all cases. 
\item The most general form of $A_i$ is 
\be 
 A_i = a G_{r0i} . 
 \ee 
Under any of the four KMS transformations, this term transforms as 
\be 
a G_{r0i} G_{a0i} \to a G_{r0i} G_{a0i} + a G_{r0i }\partial_0 G_{r0i}.
\ee
The second term is always a total derivative,
\be 
a G_{r0i }\partial_0 G_{r0i} =  \frac{1}{2}\partial_0 \alpha, \quad \alpha(g) = \int_0^g dg' a (g') .
\ee
 so this term is present in all cases with an arbitrary coefficient $a$.  
\item The most general $D_{ijkl}$ terms are 
\be
D_{ijkl} =  i ( d \,  \delta_{ik} \delta_{jl} + \tilde d  \, {\tilde \epsilon}_{ij} \tilde \epsilon_{kl} )G_{aij}G_{akl}  . 
\ee
$D_{ijkl}$ and $P_{ijkl}^{(e)}$ transform in the same way under all four KMS transformation \eqref{CaseIKMS}--\eqref{CaseIVKMS}, and from this transformation we get the constraint 
\be 
P^{(e)} = i \beta_0 D .
\ee
$P_{ijkl}^{(o)}$ consists of a single term 
\be 
P_{ijkl}^{(o)} = p \, \delta_{ik} \tilde \epsilon_{jl} ,
\ee
which  transforms differently in the different cases, and the KMS conditions for this term will be discussed below. 

\item The $E_{ijkl}$, $K_{ijklm}$, and $L_{ijklmn}$ terms are always set to zero by the KMS conditions. Either the even or the odd part of these terms will change sign under each KMS transformation and hence be set to zero.  The remaining part contains a new term which cannot in general be combined into a total derivative. 
\end{enumerate} 

Based on the above conditions, the effective Lagrangian $\sL^{(2)}$ up to weight 2 is always given by either \eqref{L2_d1} or \eqref{L2_d2}. The general form of the weight 3 part of the Lagrangian is  
\begin{align} \label{l3gen}
\sL^{(3)} = ~&  F_{ijk} G_{a0i} \partial_j G_{r0k}  \nn 
& + M_{ijk} G_{a0i} G_{ajk} +   N_{ijk} G_{ajk} \partial_0 G_{r0i}  +  O_{ijk} G_{a0i} \partial_0 G_{rjk}   
\nn
&+ U_{ijklm}  G_{aij} \partial_k G_{alm} + V_{ijklm} G_{aij} \partial_k \partial_0 G_{rlm} + W_{ijklm} \partial_0 G_{rij} \partial_k G_{alm} \nn 
 & + C_{ijklmn} G_{aij}  G_{akl} \partial_{(m} G_{r0n)}  
+Q_{ijklmn}  G_{aij} \partial_0 G_{rkl} \partial_{(m} G_{r0n)}  \nn 
 & +  R_{ijklmn}G_{aij} G_{akl} G_{amn} +  S_{ijklmn}G_{aij} G_{akl} \partial_0 G_{rmn} + T_{ijklmn}  G_{aij} \partial_0 G_{rkl} \partial_0 G_{rmn}.
\end{align} 
We will analyse the KMS conditions further in each of the four cases. 

\subsection{Invariance under $\text{KMS}_I$}

In this case, the KMS transformation is 
\begin{align} 
\text{KMS}_I: & &   \tilde{G}_{a 0 i}(-t, \vx)&=-G_{a 0 i}(x)-i \beta_{0} \partial_{0} G_{r 0 i}(x)  , & \tilde{G}_{r 0 i}(-t, \vx)&=-G_{r 0 i}(x)   , \nn
&& \tilde{G}_{a i j}(-t, \vx)&=G_{a i j}(x)+i \beta_{0} \partial_{0} G_{r i j}(x) , & \tilde{G}_{r i j}(-t, \vx)&=G_{r i j}(x) .
\end{align} 
We get the following conditions on $\sL_1$:\footnote{Note that the constraint we find on $F^{(e)}$ here, and other similar constraints on $F$ in other cases, are one possible way of ensuring the KMS conditions, but might not be the most general solution.} 
\begin{align} 
& p \text{ is unconstrained},  \\ 
& F^{(o)}= 0, \\
& F^{(e)} \text{ is a constant independent of $G_{r0i}^2$, and } F_{ijk}^{(e)} = - F_{kji}^{(e)} ,\label{f_c}  \\
&M^{(e)} =0, \\
&O^{(e)} = - N^{(e)} ,\\
& O^{(o)} =  N^{(o)} = \frac{i \beta_0}{2} M^{(o)} ,\\
& U^{(o)} = 0, \\
& V^{(o)} = - W^{(o)}, \\
& V^{(e)} = W^{(e)} = \frac{i\beta_0}{2} U^{(e)}, \\
& Q^{(o)} = i \beta_0 C^{(o)}, \\
& C^{(e)} = 0 ,\\
& Q^{(e)}_{ijklmn} = - Q^{(e)}_{klijmn} ,\label{q_c0}  \\
& S^{(e)} = \frac{3}{2} i \beta_0 R^{(e)}, \quad T^{(e)} = - \frac{\beta_0^2}{2} R^{(e)}, \\
& R^{(o)}=0, \\
&  T^{(o)} =  i \beta_0 S^{(o)}.
\end{align} 

So for this case, we end up with the following Lagrangian: 
\be 
\sL_{ \CT_-}^{(2)} = \sL_{ \rm D2}^{(2)} ,
\ee
and
\begin{align} 
\sL_{ \CT_-}^{(3)} = &~ F_{ijk}^{(e)} G_{a0i} \partial_j G_{r0k}  \nn 
& + O^{(o)}_{ijk} ( - i \frac{2}{\beta_0} G_{a0i} G_{ajk} +    G_{ajk} \partial_0 G_{r0i}  + G_{a0i} \partial_0 G_{rjk} ) \nn 
&+  O^{(e)}_{ijk} ( G_{a0i} \partial_0 G_{rjk} - G_{ajk} \partial_0 G_{r0i} ) 
\nn
&+ V_{ijklm}^{(e)} ( - i \frac{2}{\beta_0}G_{aij} \partial_k G_{alm}  +   G_{aij} \partial_k \partial_0 G_{rlm} + \partial_0 G_{rij} \partial_k G_{alm} ) \nn 
& + V^{(o)}_{ijklm} (G_{aij} \partial_k \partial_0 G_{rlm}  -  \partial_0 G_{rij} \partial_k G_{alm} )  
 \nn 
 & + C^{(o)}_{ijklmn} (G_{aij}  G_{akl} \partial_{(m} G_{r0n)} + i \beta_0   
 G_{aij} \partial_0 G_{rkl} \partial_{(m} G_{r0n)}) \nn 
& + Q_{ijklmn}^{(e)}G_{aij} \partial_0 G_{rkl} \partial_{(m} G_{r0n)} 
  \nn 
 & +  R_{ijklmn}^{(e)} (G_{aij} G_{akl} G_{amn} + \frac{3}{2} i \beta_0 G_{aij} G_{akl} \partial_0 G_{rmn} - \frac{\beta_0^2}{2}  G_{aij} \partial_0 G_{rkl} \partial_0 G_{rmn}) \nn 
 & + S^{(o)}_{ijklmn} (G_{aij} G_{akl} \partial_0 G_{rmn} +  i \beta_0 G_{aij} \partial_0 G_{rkl} \partial_0 G_{rmn}) ,
\end{align}
where we should remember the constraints \eqref{f_c} and \eqref{q_c0} on $F^{(e)}$ and $Q^{(e)}$.

\subsection{Invariance under $\text{KMS}_{II}$}

In this case, the KMS transformation is 
\begin{align} 
\text{KMS}_{II}: & &  \tilde{G}_{a 0 i}(-t, \vx)&=G_{a 0 i}(x) + i \beta_{0} \partial_{0} G_{r 0 i}(x) , & \tilde{G}_{r 0 i}(-t, \vx)&=G_{r 0 i}(x), \nn 
&& \tilde{G}_{a i j}(-t, \vx) &=- G_{a i j}(x)- i \beta_{0} \partial_{0} G_{r i j}(x)  ,   &  \tilde{G}_{r i j}(-t, \vx)&=- G_{r i j}(x).
\end{align} 
We get the following conditions: 
\begin{align} 
& p = 0, \\
& F \text{ is a constant independent of $G_{r0i}^2$, and } F_{ijk} = - F_{kji} ,
\label{f2_c}
\\
& M = 0, \\
& O = - N ,\\
 & V= W = \frac{i\beta_0}{2} U ,\\ 
 & Q = i \beta_0 C, \\
 & S = \frac{3}{2} i \beta_0 R, \quad T = -\frac{\beta_0^2}{2} R ,
\end{align} 
from which the Lagrangian is 
\be 
\sL^{(2)}_{ \CC \CT_-} = \sL^{(2)}_{\rm D1} ,
\ee
and
\begin{align} 
\sL^{(3)}_{ \CC \CT_-} =  &~F_{ijk} G_{a0i} \partial_j G_{r0k}  \nn 
&  +  O_{ijk} (G_{a0i} \partial_0 G_{rjk}- G_{ajk} \partial_0 G_{r0i}  ) 
\nn
&+ V_{ijklm} ( - i \frac{2}{\beta_0} G_{aij} \partial_k G_{alm}  +  G_{aij} \partial_k \partial_0 G_{rlm} +  \partial_0 G_{rij} \partial_k G_{alm} ) 
 \nn 
 & + C_{ijklmn} (G_{aij}  G_{akl} \partial_{(m} G_{r0n)} + i \beta_0   
  G_{aij} \partial_0 G_{rkl} \partial_{(m} G_{r0n)}) \nn 
   & +  R_{ijklmn} (G_{aij} G_{akl} G_{amn} + \frac{3}{2} i \beta_0 G_{aij} G_{akl} \partial_0 G_{rmn} - \frac{\beta_0^2}{2}  G_{aij} \partial_0 G_{rkl} \partial_0 G_{rmn}) ,
\end{align}
where we should remember the constraint \eqref{f2_c}.

\subsection{Invariance under $\text{KMS}_{III}$}

In this case, the KMS transformation is
\begin{align} 
\text{KMS}_{III}: & &\tilde{G}_{a\mu \nu} (-t, - \vx) &= G_{a\mu\nu} (x) + i \beta_0 \pt_0 G_{r \mu \nu} (x) , & \tilde{G}_{r\mu\nu} (-t, -\vx) &=  G_{r \mu \nu} (x)  .
\end{align} 
In this case, we get the conditions 
\begin{align} 
& p = 0 ,\\
& F = 0 ,\\
& O = N = \frac{i \beta_0}{2} M, \\
& U = 0, \\
& W = - V ,\\
& C = 0, \\
& Q_{ijklmn} = - Q_{klijmn}, \\
& S = \frac{3}{2} i \beta_0 R, \quad T = -\frac{\beta_0^2}{2} R .
\end{align} 
We find that  
\be
 \sL_{\CC \CP_- \CT_-}^{(2)} =  \sL_{\rm D1}^{(2)} ,
\ee
and
\begin{align} 
\sL_{\CC \CP_- \CT_-}^{(3)}  = & ~  O_{ijk} ( - i \frac{2}{\beta_0}G_{a0i} G_{ajk} +    G_{ajk} \partial_0 G_{r0i}  + G_{a0i} \partial_0 G_{rjk} )
\nn
& + V_{ijklm} (G_{aij} \partial_k \partial_0 G_{rlm}  -  \partial_0 G_{rij} \partial_k G_{alm} )  
 \nn 
& + Q_{ijklmn}G_{aij} \partial_0 G_{rkl} \partial_{(m} G_{r0n)} 
  \nn 
 & +  R_{ijklmn} (G_{aij} G_{akl} G_{amn} + \frac{3}{2} i \beta_0 G_{aij} G_{akl} \partial_0 G_{rmn} - \frac{\beta_0^2}{2}  G_{aij} \partial_0 G_{rkl} \partial_0 G_{rmn}) .
 \end{align} 

\subsection{Invariance under $\text{KMS}_{IV}$}

In this case, the KMS transformation is
\begin{align} 
\text{KMS}_{IV}: & &\tilde{G}_{a\mu \nu} (-t, -\vx) &= -G_{a\mu\nu} (x) - i \beta_0 \pt_0 G_{r \mu \nu} (x) , & \tilde{G}_{r\mu\nu} (-t, -\vx) &=  - G_{r \mu \nu} (x)  ,
\end{align} 

We get the following conditions: 
\begin{align} 
&p \text{ is unconstrained}, \\
& F^{(e)}= 0, \\
& F^{(o)} \text{ is a constant independent of $G_{r0i}^2$, and } F_{ijk}^{(o)} = - F_{kji}^{(o)}, \\
&M^{(o)} =0 ,\\
&O^{(o)} = - N^{(o)}, \\
& O^{(e)} =  N^{(e)} = \frac{i \beta_0}{2} M^{(e)} ,\\  
& U^{(e)} = 0, \\
& V^{(e)} = - W^{(e)} ,\\
& V^{(o)} = W^{(o)} = \frac{i\beta_0}{2} U^{(o)} ,\\
& Q^{(e)} = i \beta_0 C^{(e)}, \\
& C^{(o)} = 0, \\
& Q^{(o)}_{ijklmn} = - Q^{(o)}_{klijmn}, \label{q_c}  \\
& S^{(o)} = \frac{3}{2} i \beta_0 R^{(o)}, \quad T^{(o)} = - \frac{\beta_0^2}{2} R^{(o)}, \\
& R^{(e)}=0 ,\\
&  T^{(e)} =  i \beta_0 S^{(e)}.
\end{align} 
The Lagrangian is then
\be
\sL_{ \CP_- \CT_-}^{(2)} =  \sL_{ \rm D2}^{(2)} ,
\ee
and
\begin{align} 
\sL_{ \CP_- \CT_-}^{(3)} = & ~  p \delta_{ik} \tilde \epsilon_{jl} G_{aij} \partial_0 G_{rkl}  \nn 
&+ F_{ijk}^{(o)} G_{a0i} \partial_j G_{r0k}  \nn 
& + O^{(e)}_{ijk} ( - i \frac{2}{\beta_0}G_{a0i} G_{ajk} +    G_{ajk} \partial_0 G_{r0i}  + G_{a0i} \partial_0 G_{rjk} ) \nn 
& +  O^{(o)}_{ijk} ( G_{a0i} \partial_0 G_{rjk}- G_{ajk} \partial_0 G_{r0i} ) 
\nn
&+ V_{ijklm}^{(o)} (- i \frac{2}{\beta_0} G_{aij} \partial_k G_{alm}  +  G_{aij} \partial_k \partial_0 G_{rlm} +  \partial_0 G_{rij} \partial_k G_{alm} ) \nn 
& + V^{(e)}_{ijklm} (G_{aij} \partial_k \partial_0 G_{rlm}  -  \partial_0 G_{rij} \partial_k G_{alm} )  
 \nn 
 & + C^{(e)}_{ijklmn} (G_{aij}  G_{akl} \partial_{(m} G_{r0n)} + i \beta_0   
 G_{aij} \partial_0 G_{rkl} \partial_{(m} G_{r0n)}) \nn 
& + Q_{ijklmn}^{(o)}G_{aij} \partial_0 G_{rkl} \partial_{(m} G_{r0n)} 
  \nn 
 & +  R_{ijklmn}^{(o)} (G_{aij} G_{akl} G_{amn} + \frac{3}{2} i \beta_0 G_{aij} G_{akl} \partial_0 G_{rmn} - \frac{\beta_0^2}{2}  G_{aij} \partial_0 G_{rkl} \partial_0 G_{rmn}) \nn 
 & + S^{(e)}_{ijklmn} (G_{aij} G_{akl} \partial_0 G_{rmn} +  i \beta_0 G_{aij} \partial_0 G_{rkl} \partial_0 G_{rmn}) .
\end{align}

\bibliography{biblio}

\begin{thebibliography}{37}%
\makeatletter
\providecommand \@ifxundefined [1]{%
 \@ifx{#1\undefined}
}%
\providecommand \@ifnum [1]{%
 \ifnum #1\expandafter \@firstoftwo
 \else \expandafter \@secondoftwo
 \fi
}%
\providecommand \@ifx [1]{%
 \ifx #1\expandafter \@firstoftwo
 \else \expandafter \@secondoftwo
 \fi
}%
\providecommand \natexlab [1]{#1}%
\providecommand \enquote  [1]{``#1''}%
\providecommand \bibnamefont  [1]{#1}%
\providecommand \bibfnamefont [1]{#1}%
\providecommand \citenamefont [1]{#1}%
\providecommand \href@noop [0]{\@secondoftwo}%
\providecommand \href [0]{\begingroup \@sanitize@url \@href}%
\providecommand \@href[1]{\@@startlink{#1}\@@href}%
\providecommand \@@href[1]{\endgroup#1\@@endlink}%
\providecommand \@sanitize@url [0]{\catcode `\\12\catcode `\$12\catcode
  `\&12\catcode `\#12\catcode `\^12\catcode `\_12\catcode `\%12\relax}%
\providecommand \@@startlink[1]{}%
\providecommand \@@endlink[0]{}%
\providecommand \url  [0]{\begingroup\@sanitize@url \@url }%
\providecommand \@url [1]{\endgroup\@href {#1}{\urlprefix }}%
\providecommand \urlprefix  [0]{URL }%
\providecommand \Eprint [0]{\href }%
\providecommand \doibase [0]{http://dx.doi.org/}%
\providecommand \selectlanguage [0]{\@gobble}%
\providecommand \bibinfo  [0]{\@secondoftwo}%
\providecommand \bibfield  [0]{\@secondoftwo}%
\providecommand \translation [1]{[#1]}%
\providecommand \BibitemOpen [0]{}%
\providecommand \bibitemStop [0]{}%
\providecommand \bibitemNoStop [0]{.\EOS\space}%
\providecommand \EOS [0]{\spacefactor3000\relax}%
\providecommand \BibitemShut  [1]{\csname bibitem#1\endcsname}%
\let\auto@bib@innerbib\@empty
\bibitem [{\citenamefont {Vardhan}\ \emph {et~al.}(2022)\citenamefont
  {Vardhan}, \citenamefont {Grozdanov}, \citenamefont {Leutheusser},\ and\
  \citenamefont {Liu}}]{Vardhan:2022wxz}%
  \BibitemOpen
  \bibfield  {author} {\bibinfo {author} {\bibfnamefont {Shreya}\ \bibnamefont
  {Vardhan}}, \bibinfo {author} {\bibfnamefont {Sa\v{s}o}\ \bibnamefont
  {Grozdanov}}, \bibinfo {author} {\bibfnamefont {Samuel}\ \bibnamefont
  {Leutheusser}}, \ and\ \bibinfo {author} {\bibfnamefont {Hong}\ \bibnamefont
  {Liu}},\ }\bibfield  {title} {\enquote {\bibinfo {title} {{A new formulation
  of strong-field magnetohydrodynamics for neutron stars}},}\ }\href@noop {} {\
   (\bibinfo {year} {2022})},\ \Eprint {http://arxiv.org/abs/2207.01636}
  {arXiv:2207.01636 [astro-ph.HE]} \BibitemShut {NoStop}%
\bibitem [{\citenamefont {Gaiotto}\ \emph {et~al.}(2015)\citenamefont
  {Gaiotto}, \citenamefont {Kapustin}, \citenamefont {Seiberg},\ and\
  \citenamefont {Willett}}]{Gaiotto:2014kfa}%
  \BibitemOpen
  \bibfield  {author} {\bibinfo {author} {\bibfnamefont {Davide}\ \bibnamefont
  {Gaiotto}}, \bibinfo {author} {\bibfnamefont {Anton}\ \bibnamefont
  {Kapustin}}, \bibinfo {author} {\bibfnamefont {Nathan}\ \bibnamefont
  {Seiberg}}, \ and\ \bibinfo {author} {\bibfnamefont {Brian}\ \bibnamefont
  {Willett}},\ }\bibfield  {title} {\enquote {\bibinfo {title} {{Generalized
  Global Symmetries}},}\ }\href {\doibase 10.1007/JHEP02(2015)172} {\bibfield
  {journal} {\bibinfo  {journal} {JHEP}\ }\textbf {\bibinfo {volume} {02}},\
  \bibinfo {pages} {172} (\bibinfo {year} {2015})},\ \Eprint
  {http://arxiv.org/abs/1412.5148} {arXiv:1412.5148 [hep-th]} \BibitemShut
  {NoStop}%
\bibitem [{\citenamefont {Crossley}\ \emph {et~al.}(2017)\citenamefont
  {Crossley}, \citenamefont {Glorioso},\ and\ \citenamefont
  {Liu}}]{Crossley:2015evo}%
  \BibitemOpen
  \bibfield  {author} {\bibinfo {author} {\bibfnamefont {Michael}\ \bibnamefont
  {Crossley}}, \bibinfo {author} {\bibfnamefont {Paolo}\ \bibnamefont
  {Glorioso}}, \ and\ \bibinfo {author} {\bibfnamefont {Hong}\ \bibnamefont
  {Liu}},\ }\bibfield  {title} {\enquote {\bibinfo {title} {{Effective field
  theory of dissipative fluids}},}\ }\href {\doibase 10.1007/JHEP09(2017)095}
  {\bibfield  {journal} {\bibinfo  {journal} {JHEP}\ }\textbf {\bibinfo
  {volume} {09}},\ \bibinfo {pages} {095} (\bibinfo {year} {2017})},\ \Eprint
  {http://arxiv.org/abs/1511.03646} {arXiv:1511.03646 [hep-th]} \BibitemShut
  {NoStop}%
\bibitem [{\citenamefont {Glorioso}\ \emph {et~al.}(2017)\citenamefont
  {Glorioso}, \citenamefont {Crossley},\ and\ \citenamefont
  {Liu}}]{Glorioso:2017fpd}%
  \BibitemOpen
  \bibfield  {author} {\bibinfo {author} {\bibfnamefont {Paolo}\ \bibnamefont
  {Glorioso}}, \bibinfo {author} {\bibfnamefont {Michael}\ \bibnamefont
  {Crossley}}, \ and\ \bibinfo {author} {\bibfnamefont {Hong}\ \bibnamefont
  {Liu}},\ }\bibfield  {title} {\enquote {\bibinfo {title} {{Effective field
  theory of dissipative fluids (II): classical limit, dynamical KMS symmetry
  and entropy current}},}\ }\href {\doibase 10.1007/JHEP09(2017)096} {\bibfield
   {journal} {\bibinfo  {journal} {JHEP}\ }\textbf {\bibinfo {volume} {09}},\
  \bibinfo {pages} {096} (\bibinfo {year} {2017})},\ \Eprint
  {http://arxiv.org/abs/1701.07817} {arXiv:1701.07817 [hep-th]} \BibitemShut
  {NoStop}%
\bibitem [{\citenamefont {Glorioso}\ and\ \citenamefont
  {Liu}(2018)}]{Glorioso:2018wxw}%
  \BibitemOpen
  \bibfield  {author} {\bibinfo {author} {\bibfnamefont {Paolo}\ \bibnamefont
  {Glorioso}}\ and\ \bibinfo {author} {\bibfnamefont {Hong}\ \bibnamefont
  {Liu}},\ }\bibfield  {title} {\enquote {\bibinfo {title} {{Lectures on
  non-equilibrium effective field theories and fluctuating hydrodynamics}},}\
  }\href@noop {} {\  (\bibinfo {year} {2018})},\ \Eprint
  {http://arxiv.org/abs/1805.09331} {arXiv:1805.09331 [hep-th]} \BibitemShut
  {NoStop}%
\bibitem [{\citenamefont {Grozdanov}\ and\ \citenamefont
  {Polonyi}(2015)}]{Grozdanov:2013dba}%
  \BibitemOpen
  \bibfield  {author} {\bibinfo {author} {\bibfnamefont {Sa\v{s}o}\
  \bibnamefont {Grozdanov}}\ and\ \bibinfo {author} {\bibfnamefont {Janos}\
  \bibnamefont {Polonyi}},\ }\bibfield  {title} {\enquote {\bibinfo {title}
  {{Viscosity and dissipative hydrodynamics from effective field theory}},}\
  }\href {\doibase 10.1103/PhysRevD.91.105031} {\bibfield  {journal} {\bibinfo
  {journal} {Phys. Rev.}\ }\textbf {\bibinfo {volume} {D91}},\ \bibinfo {pages}
  {105031} (\bibinfo {year} {2015})},\ \Eprint {http://arxiv.org/abs/1305.3670}
  {arXiv:1305.3670 [hep-th]} \BibitemShut {NoStop}%
\bibitem [{\citenamefont {Haehl}\ \emph {et~al.}(2016)\citenamefont {Haehl},
  \citenamefont {Loganayagam},\ and\ \citenamefont
  {Rangamani}}]{Haehl:2015uoc}%
  \BibitemOpen
  \bibfield  {author} {\bibinfo {author} {\bibfnamefont {Felix~M.}\
  \bibnamefont {Haehl}}, \bibinfo {author} {\bibfnamefont {R.}~\bibnamefont
  {Loganayagam}}, \ and\ \bibinfo {author} {\bibfnamefont {M.}~\bibnamefont
  {Rangamani}},\ }\bibfield  {title} {\enquote {\bibinfo {title} {{Topological
  sigma models \& dissipative hydrodynamics}},}\ }\href {\doibase
  10.1007/JHEP04(2016)039} {\bibfield  {journal} {\bibinfo  {journal} {JHEP}\
  }\textbf {\bibinfo {volume} {04}},\ \bibinfo {pages} {039} (\bibinfo {year}
  {2016})},\ \Eprint {http://arxiv.org/abs/1511.07809} {arXiv:1511.07809
  [hep-th]} \BibitemShut {NoStop}%
\bibitem [{\citenamefont {Jensen}\ \emph {et~al.}(2018)\citenamefont {Jensen},
  \citenamefont {Pinzani-Fokeeva},\ and\ \citenamefont
  {Yarom}}]{Jensen:2017kzi}%
  \BibitemOpen
  \bibfield  {author} {\bibinfo {author} {\bibfnamefont {Kristan}\ \bibnamefont
  {Jensen}}, \bibinfo {author} {\bibfnamefont {Natalia}\ \bibnamefont
  {Pinzani-Fokeeva}}, \ and\ \bibinfo {author} {\bibfnamefont {Amos}\
  \bibnamefont {Yarom}},\ }\bibfield  {title} {\enquote {\bibinfo {title}
  {{Dissipative hydrodynamics in superspace}},}\ }\href {\doibase
  10.1007/JHEP09(2018)127} {\bibfield  {journal} {\bibinfo  {journal} {JHEP}\
  }\textbf {\bibinfo {volume} {09}},\ \bibinfo {pages} {127} (\bibinfo {year}
  {2018})},\ \Eprint {http://arxiv.org/abs/1701.07436} {arXiv:1701.07436
  [hep-th]} \BibitemShut {NoStop}%
\bibitem [{\citenamefont {Grozdanov}\ \emph {et~al.}(2017)\citenamefont
  {Grozdanov}, \citenamefont {Hofman},\ and\ \citenamefont
  {Iqbal}}]{Grozdanov:2016tdf}%
  \BibitemOpen
  \bibfield  {author} {\bibinfo {author} {\bibfnamefont {Sa\v{s}o}\
  \bibnamefont {Grozdanov}}, \bibinfo {author} {\bibfnamefont {Diego~M.}\
  \bibnamefont {Hofman}}, \ and\ \bibinfo {author} {\bibfnamefont {Nabil}\
  \bibnamefont {Iqbal}},\ }\bibfield  {title} {\enquote {\bibinfo {title}
  {{Generalized global symmetries and dissipative magnetohydrodynamics}},}\
  }\href {\doibase 10.1103/PhysRevD.95.096003} {\bibfield  {journal} {\bibinfo
  {journal} {Phys. Rev.}\ }\textbf {\bibinfo {volume} {D95}},\ \bibinfo {pages}
  {096003} (\bibinfo {year} {2017})},\ \Eprint
  {http://arxiv.org/abs/1610.07392} {arXiv:1610.07392 [hep-th]} \BibitemShut
  {NoStop}%
\bibitem [{\citenamefont {Hernandez}\ and\ \citenamefont
  {Kovtun}(2017)}]{Hernandez:2017mch}%
  \BibitemOpen
  \bibfield  {author} {\bibinfo {author} {\bibfnamefont {Juan}\ \bibnamefont
  {Hernandez}}\ and\ \bibinfo {author} {\bibfnamefont {Pavel}\ \bibnamefont
  {Kovtun}},\ }\bibfield  {title} {\enquote {\bibinfo {title} {{Relativistic
  magnetohydrodynamics}},}\ }\href {\doibase 10.1007/JHEP05(2017)001}
  {\bibfield  {journal} {\bibinfo  {journal} {JHEP}\ }\textbf {\bibinfo
  {volume} {05}},\ \bibinfo {pages} {001} (\bibinfo {year} {2017})},\ \Eprint
  {http://arxiv.org/abs/1703.08757} {arXiv:1703.08757 [hep-th]} \BibitemShut
  {NoStop}%
\bibitem [{\citenamefont {Grozdanov}\ and\ \citenamefont
  {Poovuttikul}(2019)}]{Grozdanov:2017kyl}%
  \BibitemOpen
  \bibfield  {author} {\bibinfo {author} {\bibfnamefont {Sa\v{s}o}\
  \bibnamefont {Grozdanov}}\ and\ \bibinfo {author} {\bibfnamefont {Napat}\
  \bibnamefont {Poovuttikul}},\ }\bibfield  {title} {\enquote {\bibinfo {title}
  {{Generalised global symmetries in holography: magnetohydrodynamic waves in a
  strongly interacting plasma}},}\ }\href {\doibase 10.1007/JHEP04(2019)141}
  {\bibfield  {journal} {\bibinfo  {journal} {JHEP}\ }\textbf {\bibinfo
  {volume} {04}},\ \bibinfo {pages} {141} (\bibinfo {year} {2019})},\ \Eprint
  {http://arxiv.org/abs/1707.04182} {arXiv:1707.04182 [hep-th]} \BibitemShut
  {NoStop}%
\bibitem [{\citenamefont {Grozdanov}\ \emph {et~al.}(2019)\citenamefont
  {Grozdanov}, \citenamefont {Lucas},\ and\ \citenamefont
  {Poovuttikul}}]{Grozdanov:2018fic}%
  \BibitemOpen
  \bibfield  {author} {\bibinfo {author} {\bibfnamefont {Sa\v{s}o}\
  \bibnamefont {Grozdanov}}, \bibinfo {author} {\bibfnamefont {Andrew}\
  \bibnamefont {Lucas}}, \ and\ \bibinfo {author} {\bibfnamefont {Napat}\
  \bibnamefont {Poovuttikul}},\ }\bibfield  {title} {\enquote {\bibinfo {title}
  {{Holography and hydrodynamics with weakly broken symmetries}},}\ }\href
  {\doibase 10.1103/PhysRevD.99.086012} {\bibfield  {journal} {\bibinfo
  {journal} {Phys. Rev. D}\ }\textbf {\bibinfo {volume} {99}},\ \bibinfo
  {pages} {086012} (\bibinfo {year} {2019})},\ \Eprint
  {http://arxiv.org/abs/1810.10016} {arXiv:1810.10016 [hep-th]} \BibitemShut
  {NoStop}%
\bibitem [{\citenamefont {Glorioso}\ and\ \citenamefont
  {Son}(2018)}]{Glorioso:2018kcp}%
  \BibitemOpen
  \bibfield  {author} {\bibinfo {author} {\bibfnamefont {Paolo}\ \bibnamefont
  {Glorioso}}\ and\ \bibinfo {author} {\bibfnamefont {Dam~Thanh}\ \bibnamefont
  {Son}},\ }\bibfield  {title} {\enquote {\bibinfo {title} {{Effective field
  theory of magnetohydrodynamics from generalized global symmetries}},}\
  }\href@noop {} {\  (\bibinfo {year} {2018})},\ \Eprint
  {http://arxiv.org/abs/1811.04879} {arXiv:1811.04879 [hep-th]} \BibitemShut
  {NoStop}%
\bibitem [{\citenamefont {Armas}\ and\ \citenamefont
  {Jain}(2020)}]{Armas:2018zbe}%
  \BibitemOpen
  \bibfield  {author} {\bibinfo {author} {\bibfnamefont {Jay}\ \bibnamefont
  {Armas}}\ and\ \bibinfo {author} {\bibfnamefont {Akash}\ \bibnamefont
  {Jain}},\ }\bibfield  {title} {\enquote {\bibinfo {title} {{One-form
  superfluids \& magnetohydrodynamics}},}\ }\href {\doibase
  10.1007/JHEP01(2020)041} {\bibfield  {journal} {\bibinfo  {journal} {JHEP}\
  }\textbf {\bibinfo {volume} {01}},\ \bibinfo {pages} {041} (\bibinfo {year}
  {2020})},\ \Eprint {http://arxiv.org/abs/1811.04913} {arXiv:1811.04913
  [hep-th]} \BibitemShut {NoStop}%
\bibitem [{\citenamefont {Gralla}\ and\ \citenamefont
  {Iqbal}(2019)}]{Gralla:2018kif}%
  \BibitemOpen
  \bibfield  {author} {\bibinfo {author} {\bibfnamefont {Samuel~E.}\
  \bibnamefont {Gralla}}\ and\ \bibinfo {author} {\bibfnamefont {Nabil}\
  \bibnamefont {Iqbal}},\ }\bibfield  {title} {\enquote {\bibinfo {title}
  {{Effective Field Theory of Force-Free Electrodynamics}},}\ }\href {\doibase
  10.1103/PhysRevD.99.105004} {\bibfield  {journal} {\bibinfo  {journal} {Phys.
  Rev. D}\ }\textbf {\bibinfo {volume} {99}},\ \bibinfo {pages} {105004}
  (\bibinfo {year} {2019})},\ \Eprint {http://arxiv.org/abs/1811.07438}
  {arXiv:1811.07438 [hep-th]} \BibitemShut {NoStop}%
\bibitem [{\citenamefont {Benenowski}\ and\ \citenamefont
  {Poovuttikul}(2019)}]{Benenowski:2019ule}%
  \BibitemOpen
  \bibfield  {author} {\bibinfo {author} {\bibfnamefont {Bartosz}\ \bibnamefont
  {Benenowski}}\ and\ \bibinfo {author} {\bibfnamefont {Napat}\ \bibnamefont
  {Poovuttikul}},\ }\bibfield  {title} {\enquote {\bibinfo {title}
  {{Classification of magnetohydrodynamic transport at strong magnetic
  field}},}\ }\href@noop {} {\  (\bibinfo {year} {2019})},\ \Eprint
  {http://arxiv.org/abs/1911.05554} {arXiv:1911.05554 [hep-th]} \BibitemShut
  {NoStop}%
\bibitem [{\citenamefont {Iqbal}\ and\ \citenamefont
  {Poovuttikul}(2023)}]{Iqbal:2020lrt}%
  \BibitemOpen
  \bibfield  {author} {\bibinfo {author} {\bibfnamefont {Nabil}\ \bibnamefont
  {Iqbal}}\ and\ \bibinfo {author} {\bibfnamefont {Napat}\ \bibnamefont
  {Poovuttikul}},\ }\bibfield  {title} {\enquote {\bibinfo {title} {{2-group
  global symmetries, hydrodynamics and holography}},}\ }\href {\doibase
  10.21468/SciPostPhys.15.2.063} {\bibfield  {journal} {\bibinfo  {journal}
  {SciPost Phys.}\ }\textbf {\bibinfo {volume} {15}},\ \bibinfo {pages} {063}
  (\bibinfo {year} {2023})},\ \Eprint {http://arxiv.org/abs/2010.00320}
  {arXiv:2010.00320 [hep-th]} \BibitemShut {NoStop}%
\bibitem [{\citenamefont {Landry}(2021)}]{Landry:2021kko}%
  \BibitemOpen
  \bibfield  {author} {\bibinfo {author} {\bibfnamefont {Michael~J.}\
  \bibnamefont {Landry}},\ }\bibfield  {title} {\enquote {\bibinfo {title}
  {{Higher-form and (non-)St\"uckelberg symmetries in non-equilibrium
  systems}},}\ }\href@noop {} {\  (\bibinfo {year} {2021})},\ \Eprint
  {http://arxiv.org/abs/2101.02210} {arXiv:2101.02210 [hep-th]} \BibitemShut
  {NoStop}%
\bibitem [{\citenamefont {Armas}\ and\ \citenamefont
  {Camilloni}(2022)}]{Armas:2022wvb}%
  \BibitemOpen
  \bibfield  {author} {\bibinfo {author} {\bibfnamefont {Jay}\ \bibnamefont
  {Armas}}\ and\ \bibinfo {author} {\bibfnamefont {Filippo}\ \bibnamefont
  {Camilloni}},\ }\bibfield  {title} {\enquote {\bibinfo {title} {{A stable and
  causal model of magnetohydrodynamics}},}\ }\href@noop {} {\  (\bibinfo {year}
  {2022})},\ \Eprint {http://arxiv.org/abs/2201.06847} {arXiv:2201.06847
  [hep-th]} \BibitemShut {NoStop}%
\bibitem [{\citenamefont {Das}\ \emph {et~al.}(2022)\citenamefont {Das},
  \citenamefont {Iqbal},\ and\ \citenamefont {Poovuttikul}}]{Das:2022fho}%
  \BibitemOpen
  \bibfield  {author} {\bibinfo {author} {\bibfnamefont {Arpit}\ \bibnamefont
  {Das}}, \bibinfo {author} {\bibfnamefont {Nabil}\ \bibnamefont {Iqbal}}, \
  and\ \bibinfo {author} {\bibfnamefont {Napat}\ \bibnamefont {Poovuttikul}},\
  }\bibfield  {title} {\enquote {\bibinfo {title} {{Towards an effective action
  for chiral magnetohydrodynamics}},}\ }\href@noop {} {\  (\bibinfo {year}
  {2022})},\ \Eprint {http://arxiv.org/abs/2212.09787} {arXiv:2212.09787
  [hep-th]} \BibitemShut {NoStop}%
\bibitem [{\citenamefont {Das}\ \emph {et~al.}(2023)\citenamefont {Das},
  \citenamefont {Florio}, \citenamefont {Iqbal},\ and\ \citenamefont
  {Poovuttikul}}]{Das:2023nwl}%
  \BibitemOpen
  \bibfield  {author} {\bibinfo {author} {\bibfnamefont {Arpit}\ \bibnamefont
  {Das}}, \bibinfo {author} {\bibfnamefont {Adrien}\ \bibnamefont {Florio}},
  \bibinfo {author} {\bibfnamefont {Nabil}\ \bibnamefont {Iqbal}}, \ and\
  \bibinfo {author} {\bibfnamefont {Napat}\ \bibnamefont {Poovuttikul}},\
  }\bibfield  {title} {\enquote {\bibinfo {title} {{Higher-form symmetry and
  chiral transport in real-time lattice $U(1)$ gauge theory}},}\ }\href@noop {}
  {\  (\bibinfo {year} {2023})},\ \Eprint {http://arxiv.org/abs/2309.14438}
  {arXiv:2309.14438 [hep-th]} \BibitemShut {NoStop}%
\bibitem [{\citenamefont {Frangi}\ and\ \citenamefont
  {Grozdanov}(2024)}]{Frangi:2024mer}%
  \BibitemOpen
  \bibfield  {author} {\bibinfo {author} {\bibfnamefont {Giorgio}\ \bibnamefont
  {Frangi}}\ and\ \bibinfo {author} {\bibfnamefont {Sa\v{s}o}\ \bibnamefont
  {Grozdanov}},\ }\bibfield  {title} {\enquote {\bibinfo {title} {{Quantum
  origin of Ohm's reciprocity relation and its violation: conductivity as
  inverse resistivity}},}\ }\href@noop {} {\  (\bibinfo {year} {2024})},\
  \Eprint {http://arxiv.org/abs/2406.16123} {arXiv:2406.16123 [hep-th]}
  \BibitemShut {NoStop}%
\bibitem [{\citenamefont {Frangi}(2024)}]{Frangi:2024enh}%
  \BibitemOpen
  \bibfield  {author} {\bibinfo {author} {\bibfnamefont {Giorgio}\ \bibnamefont
  {Frangi}},\ }\bibfield  {title} {\enquote {\bibinfo {title} {{Geometrisation
  of Ohm's reciprocity relation in a holographic plasma}},}\ }\href@noop {} {\
  (\bibinfo {year} {2024})},\ \Eprint {http://arxiv.org/abs/2406.16124}
  {arXiv:2406.16124 [hep-th]} \BibitemShut {NoStop}%
\bibitem [{\citenamefont {Landry}\ and\ \citenamefont
  {Liu}(2022)}]{Landry:2022nog}%
  \BibitemOpen
  \bibfield  {author} {\bibinfo {author} {\bibfnamefont {Michael~J.}\
  \bibnamefont {Landry}}\ and\ \bibinfo {author} {\bibfnamefont {Hong}\
  \bibnamefont {Liu}},\ }\bibfield  {title} {\enquote {\bibinfo {title} {{A
  systematic formulation of chiral anomalous magnetohydrodynamics}},}\
  }\href@noop {} {\  (\bibinfo {year} {2022})},\ \Eprint
  {http://arxiv.org/abs/2212.09757} {arXiv:2212.09757 [hep-ph]} \BibitemShut
  {NoStop}%
\bibitem [{\citenamefont {{Caldarelli}}\ \emph {et~al.}(2011)\citenamefont
  {{Caldarelli}}, \citenamefont {{Emparan}},\ and\ \citenamefont {{van
  Pol}}}]{emparan}%
  \BibitemOpen
  \bibfield  {author} {\bibinfo {author} {\bibfnamefont {Marco~M.}\
  \bibnamefont {{Caldarelli}}}, \bibinfo {author} {\bibfnamefont {Roberto}\
  \bibnamefont {{Emparan}}}, \ and\ \bibinfo {author} {\bibfnamefont {Bert}\
  \bibnamefont {{van Pol}}},\ }\bibfield  {title} {\enquote {\bibinfo {title}
  {{Higher-dimensional rotating charged black holes}},}\ }\href {\doibase
  10.1007/JHEP04(2011)013} {\bibfield  {journal} {\bibinfo  {journal} {Journal
  of High Energy Physics}\ }\textbf {\bibinfo {volume} {2011}},\ \bibinfo {eid}
  {13} (\bibinfo {year} {2011})},\ \Eprint {http://arxiv.org/abs/1012.4517}
  {arXiv:1012.4517 [hep-th]} \BibitemShut {NoStop}%
\bibitem [{\citenamefont {{Emparan}}\ \emph {et~al.}(2011)\citenamefont
  {{Emparan}}, \citenamefont {{Harmark}}, \citenamefont {{Niarchos}},\ and\
  \citenamefont {{Obers}}}]{emparan_2}%
  \BibitemOpen
  \bibfield  {author} {\bibinfo {author} {\bibfnamefont {Roberto}\ \bibnamefont
  {{Emparan}}}, \bibinfo {author} {\bibfnamefont {Troels}\ \bibnamefont
  {{Harmark}}}, \bibinfo {author} {\bibfnamefont {Vasilis}\ \bibnamefont
  {{Niarchos}}}, \ and\ \bibinfo {author} {\bibfnamefont {Niels~A.}\
  \bibnamefont {{Obers}}},\ }\bibfield  {title} {\enquote {\bibinfo {title}
  {{Blackfolds in supergravity and string theory}},}\ }\href {\doibase
  10.1007/JHEP08(2011)154} {\bibfield  {journal} {\bibinfo  {journal} {Journal
  of High Energy Physics}\ }\textbf {\bibinfo {volume} {2011}},\ \bibinfo {eid}
  {154} (\bibinfo {year} {2011})},\ \Eprint {http://arxiv.org/abs/1106.4428}
  {arXiv:1106.4428 [hep-th]} \BibitemShut {NoStop}%
\bibitem [{\citenamefont {{Schubring}}(2015)}]{daniel}%
  \BibitemOpen
  \bibfield  {author} {\bibinfo {author} {\bibfnamefont {Daniel}\ \bibnamefont
  {{Schubring}}},\ }\bibfield  {title} {\enquote {\bibinfo {title}
  {{Dissipative string fluids}},}\ }\href {\doibase 10.1103/PhysRevD.91.043518}
  {\bibfield  {journal} {\bibinfo  {journal} {\prd}\ }\textbf {\bibinfo
  {volume} {91}},\ \bibinfo {eid} {043518} (\bibinfo {year} {2015})},\ \Eprint
  {http://arxiv.org/abs/1412.3135} {arXiv:1412.3135 [hep-th]} \BibitemShut
  {NoStop}%
\bibitem [{\citenamefont {Glorioso}\ \emph {et~al.}(2019)\citenamefont
  {Glorioso}, \citenamefont {Liu},\ and\ \citenamefont
  {Rajagopal}}]{Glorioso:2017lcn}%
  \BibitemOpen
  \bibfield  {author} {\bibinfo {author} {\bibfnamefont {Paolo}\ \bibnamefont
  {Glorioso}}, \bibinfo {author} {\bibfnamefont {Hong}\ \bibnamefont {Liu}}, \
  and\ \bibinfo {author} {\bibfnamefont {Srivatsan}\ \bibnamefont
  {Rajagopal}},\ }\bibfield  {title} {\enquote {\bibinfo {title} {{Global
  Anomalies, Discrete Symmetries, and Hydrodynamic Effective Actions}},}\
  }\href {\doibase 10.1007/JHEP01(2019)043} {\bibfield  {journal} {\bibinfo
  {journal} {JHEP}\ }\textbf {\bibinfo {volume} {01}},\ \bibinfo {pages} {043}
  (\bibinfo {year} {2019})},\ \Eprint {http://arxiv.org/abs/1710.03768}
  {arXiv:1710.03768 [hep-th]} \BibitemShut {NoStop}%
\bibitem [{\citenamefont {{Goldreich}}\ and\ \citenamefont
  {{Reisenegger}}(1992)}]{reis}%
  \BibitemOpen
  \bibfield  {author} {\bibinfo {author} {\bibfnamefont {Peter}\ \bibnamefont
  {{Goldreich}}}\ and\ \bibinfo {author} {\bibfnamefont {Andreas}\ \bibnamefont
  {{Reisenegger}}},\ }\bibfield  {title} {\enquote {\bibinfo {title} {{Magnetic
  Field Decay in Isolated Neutron Stars}},}\ }\href {\doibase 10.1086/171646}
  {\bibfield  {journal} {\bibinfo  {journal} {\apj}\ }\textbf {\bibinfo
  {volume} {395}},\ \bibinfo {pages} {250} (\bibinfo {year}
  {1992})}\BibitemShut {NoStop}%
\bibitem [{\citenamefont {{Armitage}}\ \emph {et~al.}(2018)\citenamefont
  {{Armitage}}, \citenamefont {{Mele}},\ and\ \citenamefont
  {{Vishwanath}}}]{weyl_review}%
  \BibitemOpen
  \bibfield  {author} {\bibinfo {author} {\bibfnamefont {N.~P.}\ \bibnamefont
  {{Armitage}}}, \bibinfo {author} {\bibfnamefont {E.~J.}\ \bibnamefont
  {{Mele}}}, \ and\ \bibinfo {author} {\bibfnamefont {Ashvin}\ \bibnamefont
  {{Vishwanath}}},\ }\bibfield  {title} {\enquote {\bibinfo {title} {{Weyl and
  Dirac semimetals in three-dimensional solids}},}\ }\href {\doibase
  10.1103/RevModPhys.90.015001} {\bibfield  {journal} {\bibinfo  {journal}
  {Reviews of Modern Physics}\ }\textbf {\bibinfo {volume} {90}},\ \bibinfo
  {eid} {015001} (\bibinfo {year} {2018})},\ \Eprint
  {http://arxiv.org/abs/1705.01111} {arXiv:1705.01111 [cond-mat.str-el]}
  \BibitemShut {NoStop}%
\bibitem [{\citenamefont {{Iqbal}}\ and\ \citenamefont
  {{Poovuttikul}}(2023)}]{2group}%
  \BibitemOpen
  \bibfield  {author} {\bibinfo {author} {\bibfnamefont {Nabil}\ \bibnamefont
  {{Iqbal}}}\ and\ \bibinfo {author} {\bibfnamefont {Nick}\ \bibnamefont
  {{Poovuttikul}}},\ }\bibfield  {title} {\enquote {\bibinfo {title} {{2-group
  global symmetries, hydrodynamics and holography}},}\ }\href {\doibase
  10.21468/SciPostPhys.15.2.063} {\bibfield  {journal} {\bibinfo  {journal}
  {SciPost Physics}\ }\textbf {\bibinfo {volume} {15}},\ \bibinfo {eid} {063}
  (\bibinfo {year} {2023})},\ \Eprint {http://arxiv.org/abs/2010.00320}
  {arXiv:2010.00320 [hep-th]} \BibitemShut {NoStop}%
\bibitem [{\citenamefont {Glorioso}\ and\ \citenamefont
  {Liu}(2016)}]{Glorioso:2016gsa}%
  \BibitemOpen
  \bibfield  {author} {\bibinfo {author} {\bibfnamefont {Paolo}\ \bibnamefont
  {Glorioso}}\ and\ \bibinfo {author} {\bibfnamefont {Hong}\ \bibnamefont
  {Liu}},\ }\bibfield  {title} {\enquote {\bibinfo {title} {{The second law of
  thermodynamics from symmetry and unitarity}},}\ }\href@noop {} {\  (\bibinfo
  {year} {2016})},\ \Eprint {http://arxiv.org/abs/1612.07705} {arXiv:1612.07705
  [hep-th]} \BibitemShut {NoStop}%
\bibitem [{\citenamefont {{Banerjee}}\ \emph {et~al.}(2024)\citenamefont
  {{Banerjee}}, \citenamefont {{Moessner}},\ and\ \citenamefont
  {{Sur{\'o}wka}}}]{monopole_1}%
  \BibitemOpen
  \bibfield  {author} {\bibinfo {author} {\bibfnamefont {Debarghya}\
  \bibnamefont {{Banerjee}}}, \bibinfo {author} {\bibfnamefont {Roderich}\
  \bibnamefont {{Moessner}}}, \ and\ \bibinfo {author} {\bibfnamefont {Piotr}\
  \bibnamefont {{Sur{\'o}wka}}},\ }\bibfield  {title} {\enquote {\bibinfo
  {title} {{Monopole magnetohydrodynamics on a plane: Magnetosonic waves and
  dynamo instability}},}\ }\href {\doibase 10.48550/arXiv.2406.09562}
  {\bibfield  {journal} {\bibinfo  {journal} {arXiv e-prints}\ ,\ \bibinfo
  {eid} {arXiv:2406.09562}} (\bibinfo {year} {2024})},\ \Eprint
  {http://arxiv.org/abs/2406.09562} {arXiv:2406.09562 [cond-mat.str-el]}
  \BibitemShut {NoStop}%
\bibitem [{\citenamefont {{Armas}}\ and\ \citenamefont
  {{Jain}}(2024)}]{monopole_2}%
  \BibitemOpen
  \bibfield  {author} {\bibinfo {author} {\bibfnamefont {Jay}\ \bibnamefont
  {{Armas}}}\ and\ \bibinfo {author} {\bibfnamefont {Akash}\ \bibnamefont
  {{Jain}}},\ }\bibfield  {title} {\enquote {\bibinfo {title} {{Approximate
  higher-form symmetries, topological defects, and dynamical phase
  transitions}},}\ }\href {\doibase 10.1103/PhysRevD.109.045019} {\bibfield
  {journal} {\bibinfo  {journal} {\prd}\ }\textbf {\bibinfo {volume} {109}},\
  \bibinfo {eid} {045019} (\bibinfo {year} {2024})},\ \Eprint
  {http://arxiv.org/abs/2301.09628} {arXiv:2301.09628 [hep-th]} \BibitemShut
  {NoStop}%
\bibitem [{\citenamefont {Lake}(2018)}]{Lake:2018dqm}%
  \BibitemOpen
  \bibfield  {author} {\bibinfo {author} {\bibfnamefont {Ethan}\ \bibnamefont
  {Lake}},\ }\bibfield  {title} {\enquote {\bibinfo {title} {{Higher-form
  symmetries and spontaneous symmetry breaking}},}\ }\href@noop {} {\
  (\bibinfo {year} {2018})},\ \Eprint {http://arxiv.org/abs/1802.07747}
  {arXiv:1802.07747 [hep-th]} \BibitemShut {NoStop}%
\bibitem [{\citenamefont {Hofman}\ and\ \citenamefont
  {Iqbal}(2019)}]{Hofman:2018lfz}%
  \BibitemOpen
  \bibfield  {author} {\bibinfo {author} {\bibfnamefont {Diego~M.}\
  \bibnamefont {Hofman}}\ and\ \bibinfo {author} {\bibfnamefont {Nabil}\
  \bibnamefont {Iqbal}},\ }\bibfield  {title} {\enquote {\bibinfo {title}
  {{Goldstone modes and photonization for higher form symmetries}},}\ }\href
  {\doibase 10.21468/SciPostPhys.6.1.006} {\bibfield  {journal} {\bibinfo
  {journal} {SciPost Phys.}\ }\textbf {\bibinfo {volume} {6}},\ \bibinfo
  {pages} {006} (\bibinfo {year} {2019})},\ \Eprint
  {http://arxiv.org/abs/1802.09512} {arXiv:1802.09512 [hep-th]} \BibitemShut
  {NoStop}%
\bibitem [{\citenamefont {{Iqbal}}\ and\ \citenamefont
  {{McGreevy}}(2022)}]{iqbal_mcgreevy}%
  \BibitemOpen
  \bibfield  {author} {\bibinfo {author} {\bibfnamefont {Nabil}\ \bibnamefont
  {{Iqbal}}}\ and\ \bibinfo {author} {\bibfnamefont {John}\ \bibnamefont
  {{McGreevy}}},\ }\bibfield  {title} {\enquote {\bibinfo {title} {{Mean string
  field theory: Landau-Ginzburg theory for 1-form symmetries}},}\ }\href
  {\doibase 10.21468/SciPostPhys.13.5.114} {\bibfield  {journal} {\bibinfo
  {journal} {SciPost Physics}\ }\textbf {\bibinfo {volume} {13}},\ \bibinfo
  {eid} {114} (\bibinfo {year} {2022})},\ \Eprint
  {http://arxiv.org/abs/2106.12610} {arXiv:2106.12610 [hep-th]} \BibitemShut
  {NoStop}%
\end{thebibliography}%
\end{document}